\documentclass[aps,prl,twocolumn,groupedaddress,longbibliography,letterpaper]{revtex4-1}

\usepackage{amssymb}
\usepackage{amsmath}
\usepackage{graphicx}
\usepackage{url}
\usepackage{hyperref}
\usepackage{lipsum}
\usepackage{xcolor}

\newcommand\ket[1]{\ensuremath{|#1\rangle}}

\usepackage[utf8]{inputenc}
\DeclareUnicodeCharacter{2009}{ }

\newcommand{\beginsupplement}{%

        \setcounter{table}{0}

        \renewcommand{\thetable}{S\arabic{table}}%

        \setcounter{figure}{0}

        \renewcommand{\thefigure}{S\arabic{figure}}%

        \setcounter{equation}{0}

        \renewcommand{\theequation}{S\arabic{equation}}%

     }

\newcommand\unit[1]{\ensuremath{\ \mathrm{#1}}}

\newcommand{\Deltak}{\Delta \mathbf{k}}

\newcommand{\deltaRaman}{\delta_{\mathrm{R}}}
\newcommand{\deltav}{\delta_{\mathbf{v}}}

\newcommand{\deltaR}{\delta_{\mathrm{R}}}
\newcommand{\OmegaR}{\Omega_{\mathrm{R}}}

\newcommand{\Rb}{^{87}\mathrm{Rb}}
\newcommand{\sigmaminus}{\sigma^-}

\newcommand{\mum}{\unit{\mu{}\mathrm{m}}}
\newcommand{\omegabar}{\bar{\omega}}
\newcommand{\kB}{k_\mathrm{B}}

\newcommand{\Gammasc}{\Gamma_{\mathrm{sc}}}
\newcommand{\Erecoil}{E_{\mathrm{r}}}
\newcommand{\Trecoil}{T_{\mathrm{r}}}
\newcommand{\lambdadB}{\lambda_{\mathrm{dB}}}

\begin{document}

\title{Direct laser cooling to Bose-Einstein condensation in a dipole trap}

\author{Alban Urvoy}
\thanks{These two authors contributed equally to this work.}
\author{Zachary Vendeiro}
\thanks{These two authors contributed equally to this work.}
\author{Joshua Ramette}
\author{Albert Adiyatullin}
\author{Vladan Vuleti\'c}
\email{vuletic@mit.edu}
\affiliation{Department of Physics, MIT-Harvard Center for Ultracold Atoms and Research Laboratory of Electronics, Massachusetts Institute of Technology, Cambridge, Massachusetts 02139, USA}

\date{\today}

\begin{abstract}
We present a method for producing three-dimensional Bose-Einstein condensates using \emph{only} laser cooling. 
The phase transition to condensation is crossed with $2.5 {\times} 10^{4}$ $\Rb$ atoms at a temperature of $T_{\mathrm{c}} = 0.6 \unit{\mu K}$ after $1.4\unit{s}$ of cooling. 
Atoms are trapped in a crossed optical dipole trap and cooled using Raman cooling with far-off-resonant optical pumping light to reduce atom loss and heating. 
The achieved temperatures are well below the effective recoil temperature.
We find that during the final cooling stage at atomic densities above $10^{14} \unit{cm^{-3}}$, careful tuning of trap depth and optical-pumping rate is necessary to evade heating and loss mechanisms. 
The method may enable the fast production of quantum degenerate gases in a variety of systems including fermions.
\end{abstract}

\maketitle

\label{intro}

Quantum degenerate gases provide a flexible platform with applications ranging from quantum simulations of many-body interacting systems~\cite{Bloch2012} to precision measurements~\cite{Becker2018}.
The standard technique for achieving quantum degeneracy is laser cooling followed by evaporative cooling~\cite{Ketterle1996a} in magnetic~\cite{Anderson1995,Davis1995,DeMarco1999} or optical traps~\cite{Barrett2001}.
Evaporation is a powerful tool, but its performance depends strongly on atomic collisional properties and it requires removal of most of the initially trapped atoms.
Laser cooling gases to degeneracy could alleviate those issues, but it has proven difficult to implement.

The elusiveness of laser cooling to Bose-Einstein condensation (BEC)~\cite{Anderson1995, Davis1995} for more than two decades~\cite{Aspect1988,Lee1996,Kerman2000,Han2000} can be understood as follows: optical cooling requires spontaneous photon scattering that moves entropy from the atomic system into the light field.
Such scattering sets a natural atomic temperature scale $\Trecoil$ associated with the recoil momentum from a single photon of wavelength $2 \pi \lambdabar$ and an associated atomic thermal de Broglie wavelength \mbox{$\lambdadB = \sqrt{2 \pi \hbar^2 / (m \kB \Trecoil)} \sim \lambdabar$}.
Here $\hbar$ is the reduced Planck constant, $m$ the atomic mass, and $\kB$ the Boltzmann constant.
BEC must then be achieved at relatively high critical atomic density $n_\mathrm{c} \sim \lambdadB^{-3} \sim \lambdabar^{-3}$, where inelastic collisions result in heating and trap loss.
In particular, light-induced collisional loss becomes severe when $n \gtrsim \lambdabar^{-3}$~\cite{Burnett1996,Ido2000}.

For strontium atoms, where cooling on a spectrally narrow transition is available, a strongly-inhomogeneous trapping potential has been used to cool a lower-density thermal bath while decoupling the emerging condensate from the cooling light~\cite{Stellmer2013}.
Very recently, based on similar principles to the ones presented here, an array of small, nearly one-dimensional condensates has been prepared using degenerate Raman sideband cooling~\cite{Hamann1998,Vuletic1998,Han2000} of $\Rb$ atoms in a two-dimensional optical lattice~\cite{Hu2017}.

\begin{figure}[b]
\includegraphics[scale=1]{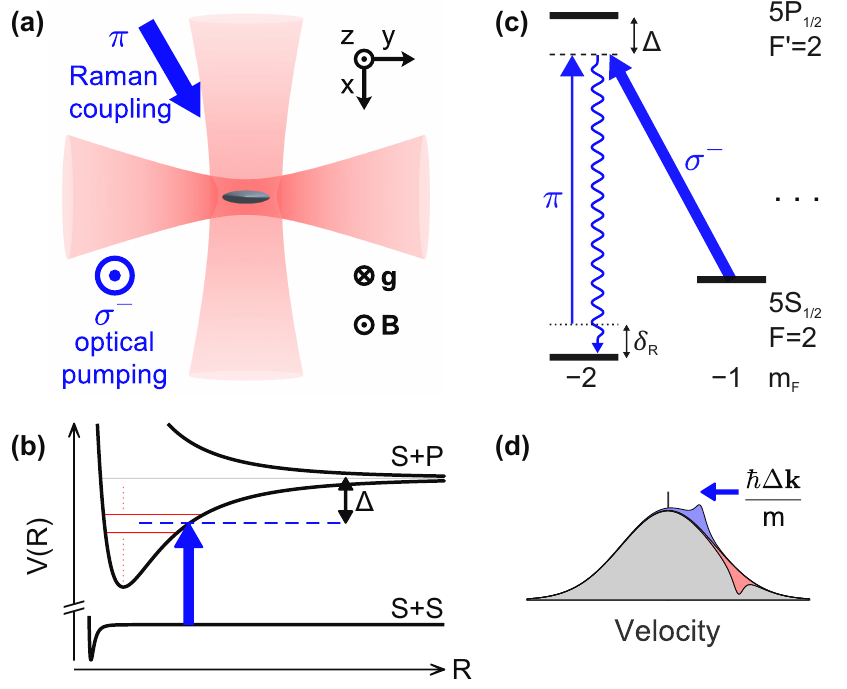}%
\caption{\label{fig1}
(a) Geometry of the experimental setup with 795nm optical pumping and Raman coupling beams, and 1064nm trapping beams.
(b) Molecular potentials. Light-assisted collisions are suppressed if the detuning from atomic resonance $\Delta$ is chosen to be far from photo-association resonances (solid red horizontal lines).
(c) Partial atomic level scheme. The Raman transition is resonant for atoms with a two-photon Doppler shift $\deltaR$.
(d) Velocity distribution of the atoms along the two-photon momentum $\hbar (\Deltak)$. A Raman transition reduces the velocity of atoms in the velocity class $\deltaR / |\Deltak|$ by $\hbar (\Deltak)/m$.
}
\end{figure}

In this Letter, we demonstrate Raman cooling~\cite{Kasevich1992,Reichel1995,Lee1996,Perrin1999,Boyer2004} of an ensemble of $\Rb$ atoms into the quantum degenerate regime, 
without evaporative cooling.
Starting with up to $1 {\times} 10^5$ atoms in an optical dipole trap, the transition to BEC is reached with up to $2.5 {\times} 10^4$ atoms within a cooling time of ${\sim} 1 \unit{s}$. 
As discussed in detail below, the essential components of our technique are 
(i) the use of carefully far-detuned cooling light to reduce atom loss and heating at high atomic densities~\cite{Burnett1996,Hijmans1996,Ruhrig2015}, 
(ii) a reduced optical pumping rate during the final stage to avoid heating by photon reabsorption (\emph{festina lente} regime~\cite{Cirac1996,Castin1998,Santos2000}), 
(iii) a final cooling of atoms in the high-energy wings of the thermal velocity distribution to achieve sub-recoil cooling, and 
(iv) careful control of the final trap depth to reduce heating induced by inelastic three-body collisions.

Raman cooling of the optically trapped atoms is a two-step process where kinetic energy is first removed via a stimulated two-photon Raman transition that simultaneously changes the internal atomic state. 
Subsequently, entropy is removed in an optical pumping process that restores the original atomic state via the spontaneous scattering of a photon [see Fig.~\ref{fig1}(c)-(d)].
The optical pumping into the state $\ket{5\mathrm{S}_{1/2}; F{=}2,m_F{=}-2}$ along the $z$-axis is performed with $\sigmaminus$-polarized light.
We reduce light-induced loss by using optical pumping light with large negative detuning \mbox{$\Delta/(2\pi) = -4.33 \unit{GHz}$} from the $\ket{5\mathrm{S}_{1/2}; F{=}2}$ to $\ket{5\mathrm{P}_{1/2}; F^\prime{=}2}$ transition of the $\mathrm{D_1}$-line, choosing a detuning far from molecular resonances [see Fig.~\ref{fig1}(b-c) and \cite{SuppMat}]. 
The far-detuned $\sigmaminus$-polarized beam and a $\pi$-polarized beam of similar detuning which propagates in the $x-y$ plane [see Fig.~\ref{fig1}(a)], drive the stimulated Raman transition to the state $\ket{5\mathrm{S}_{1/2}; 2,-1}$, simultaneously changing the atomic momentum by the two-photon recoil $\hbar (\Deltak)$~\cite{SuppMat}. 
To cool all three directions, we choose $\hbar (\Deltak)$ to have a non-zero projection along any trap eigenaxis.

\label{methods}

\begin{figure}[t]
\includegraphics[width=8.6cm]{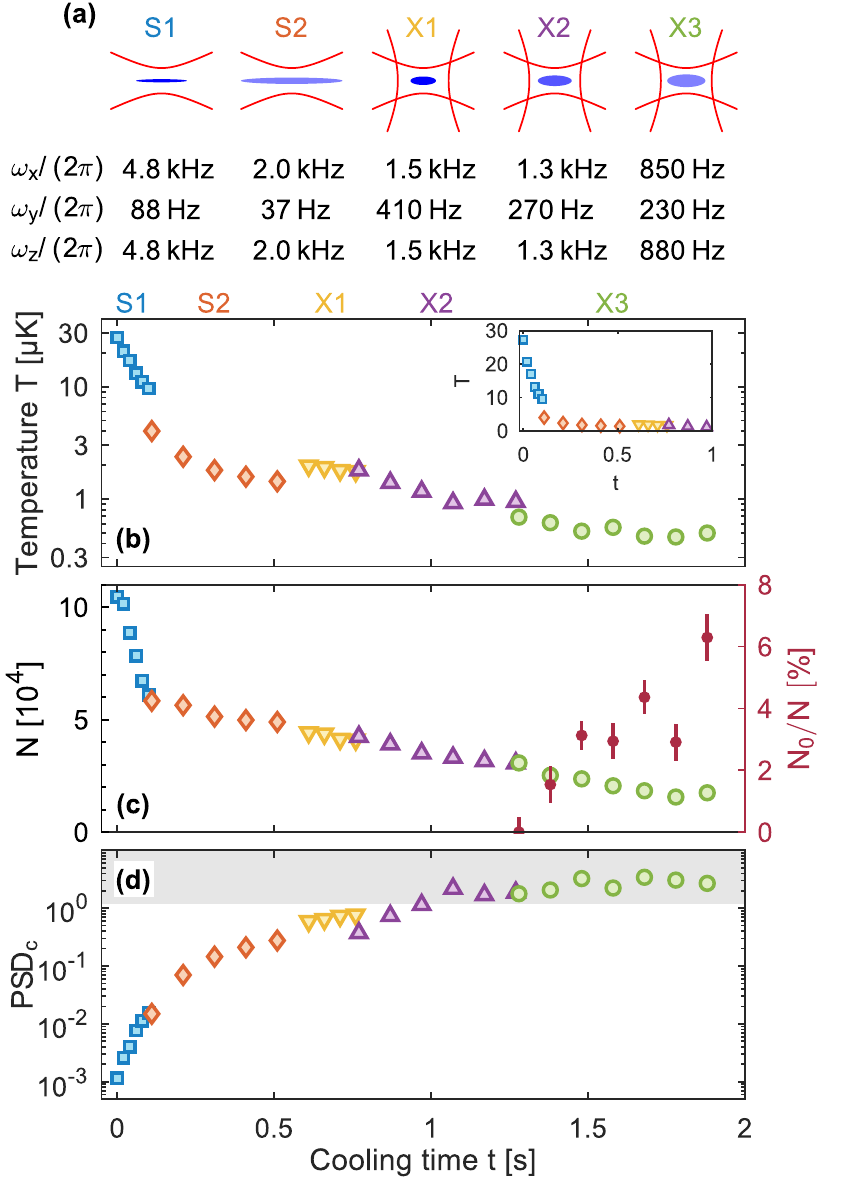}%
\caption{\label{CoolingCurves}
(a)
Schematic of the trapping potential during the cooling sequence, along with the values of the trapping frequencies for each cooling stage.
(b)
Atomic temperature $T$ as a function of cooling time $t$. Discrete jumps are caused by changes of the trapping potential between the cooling stages.
Inset: Temperature on a linear scale.
(c)
Atom number (open symbols) and condensate fraction $N_0/N$ (solid circles) during the cooling sequence.
(d)
Classical phase-space density $\mathrm{PSD_c}$ (see main text) as a function of cooling time $t$.
The grey shaded area denotes the quantum degenerate region.
Subfigures (b-d) are all plotted along the same time axis.
}
\end{figure}

Cooling is performed in several stages to allow optimization of the cooling as the atomic temperature and density change.
Within each stage, the trapping beam powers, optical pumping rate $\Gammasc$, Raman coupling Rabi frequency $\OmegaR$, and Raman resonance detuning $\deltaRaman$ are held constant.
The first two cooling stages, S1 and S2, are performed in a single-beam optical dipole trap (sODT), after which the atoms are transferred to a crossed optical dipole trap (xODT), where we perform three more cooling stages [X1 to X3, see Fig.~\ref{CoolingCurves}(a)]. 
During each stage, we characterize the cooling performance using time-of-flight absorption imaging, extracting the atom number $N$ and temperature $T$. 
(For the final cooling stage close to the BEC threshhold, we exclude the central part of the time-of-flight image from the temperature fit.) 
We quantify the cooling performance by the classical phase space density $\mathrm{PSD_{c}} = N  \left( \hbar \omegabar/(\kB T) \right)^3$, where $\omegabar=(\omega_x \omega_y \omega_z)^{1/3}$ is the geometric mean of the three trapping frequencies.
Far from degeneracy, i.e. for a classical gas, $\mathrm{PSD_c}$ is equal to the true PSD at the center of the cloud,  $\mathrm{PSD}=n(0) \lambdadB^3$.
The parameters of each stage are optimized to yield the highest $\mathrm{PSD_c}$ at the end of the stage~\cite{SuppMat}.

\label{results}

\begin{figure}[t]
\includegraphics[scale=1]{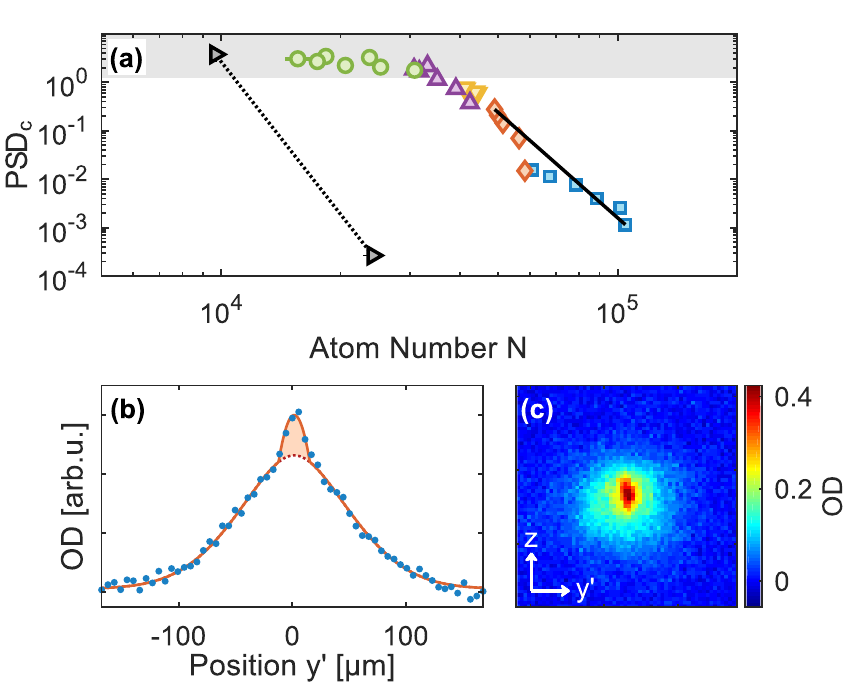}%
\caption{\label{BEC}
(a)
Classical phase-space density $\mathrm{PSD_c}$ as a function of remaining atom number $N$.
The cooling is very efficient until $\mathrm{PSD_c} {\sim} 1$ is reached.
The black symbols denote the performance of the same sequence with the initial atom number reduced by a factor 5.
The final atom number is only reduced by a factor 2, indicating a density limit in the cooling.
The solid (dashed) black line indicates the S1-S2 (S1-X3) path with an efficiency $\gamma=7.2$ ($\gamma=11$) for each case.
(b) Line optical density (dots) of the cloud in time-of-flight along the \mbox{$y^{\prime}$-direction} (slightly rotated from the $y$-direction in the $x-y$ plane, see SM~\cite{SuppMat}). 
The data is taken after $1.6 \unit{s}$ of cooling and fitted with a $g_{5{/}2}$ Bose distribution with a Thomas-Fermi distribution superimposed (orange line).
The shaded area indicates the condensed fraction.
(c)
False color image of the same cloud (before integration along the vertical direction), showing the anisotropic expansion of the condensed fraction in the center.
}
\end{figure}

\begin{figure*}[ht]
\includegraphics[scale=1]{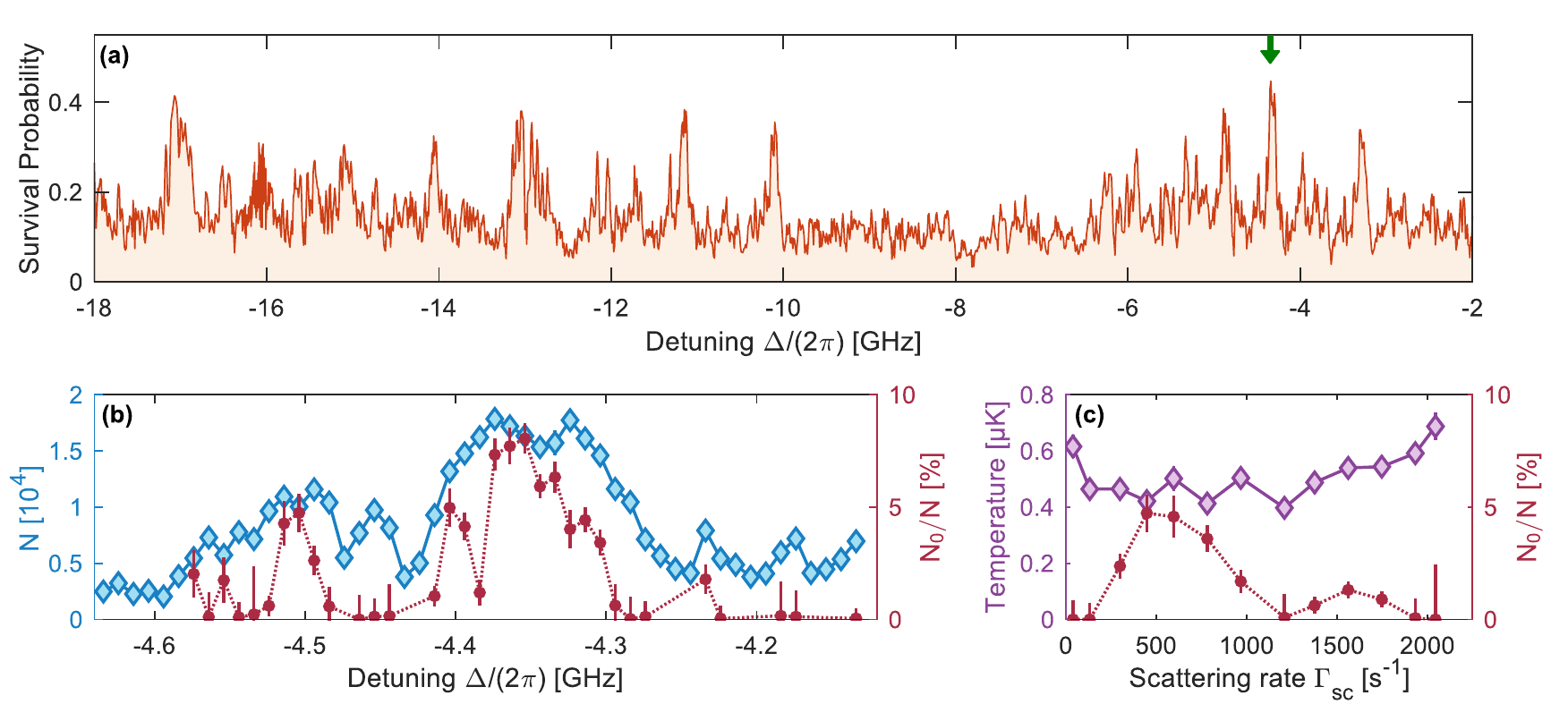}%
\caption{\label{OptimizationScans}
(a)
Photoassociation loss spectrum.
Survival probability of trapped atoms as a function of the detuning $\Delta$ of the optical pumping beam, when scattering ${\sim}100$ photons.
In substantial portions of the spectrum, the atomic loss is large, due to photoassociation resonances, whereas the peaks correspond to gaps in the photoassociation spectrum away from resonances.
The green arrow indicates the detuning used for the data in Fig.~\ref{CoolingCurves} and~\ref{BEC}.
(b)
Performance of the full cooling sequence as a function of optical pumping detuning $\Delta$ near a locally optimal detuning.
A condensate fraction $N_0/N$ is visible only when the losses are limited.
To keep the Raman resonance detuning $\deltaRaman$ constant between data points, the magnetic field is adjusted to compensate the change in light shift associated with varying $\Delta$.
(c)
Temperature and condensate fraction as a function of the scattering rate $\Gamma_{\mathrm{sc}}$ in the final cooling stage.
Here, the intensity of the $\pi$-poarized beam is adjusted to keep the Raman coupling strength constant between data points, and $\deltaRaman$ is adjusted to optimize the cooling performance for each data point.
}
\end{figure*}

We load \mbox{$1{\times}10^5 \unit{atoms}$} from a magneto-optical trap into the sODT propagating along the $y$-direction with a $10 \mum$ waist [Fig.~\ref{fig1}(a)]. 
After cooling in stage S1 (see Fig.~\ref{CoolingCurves}), the trap power and vibration frequencies are reduced, thereby lowering the density and therefore the loss in stage S2.
For all stages, we verify that the trap remains sufficiently deep to keep evaporative cooling negligible.

During stages S1 and S2 we perform fast cooling over $500 \unit{ms}$ from $30 \unit{\mu K}$ down to $1.5 \unit{\mu K}$, and up to $\mathrm{PSD_c}$ just below unity.
The larger loss rate during S1 relative to the other stages [see Fig.~\ref{CoolingCurves}(c)] occurs because the Raman cooling cycle, and hence light-induced collision rate, is faster.
The initial cooling at high temperatures $T{\approx 30} \unit{\mu K}$ and densities \mbox{$n < 7 {\times} 10^{13} \unit{cm^{-3}}$} is quite efficient, with a logarithmic slope of \mbox{$\gamma = - d (\ln{\mathrm{PSD_c}})/d(\ln{N}) =7.2$} [see Fig.~\ref{BEC}(a)]. 

After stage S2, the ensemble is sufficiently cold so that it can be efficiently transferred in the xODT by ramping up the second trapping beam ($18 \mum$ waist, propagating along $x$), and applying a short initial cooling (stage X1). 
In a similar fashion to the S1-S2 sequence, we reduce the confinement of the xODT and cool further during X2, after which the ensemble is at the threshold to condensation. 
No condensate appears in X2 despite $\mathrm{PSD_c}$ reaching the ideal-gas value of 1.2~\cite{Dalfovo1999}, which we attribute to a combination of the finite size effect~\cite{Ketterle1996}, the interaction shift~\cite{Smith2011} and small calibration errors. 
After further reduction of the xODT, we are able to cross the BEC transition during X3, as shown by the appearance of a condensed fraction in the velocity distribution [see Fig.~\ref{BEC}(b)].
The onset of BEC is further confirmed by the anisotropic expansion of the central part of the cloud due to trap confinement anisotropy [see Fig.~\ref{BEC}(c)].

In order to achieve BEC, the trap depth during X3 must not be too large. 
For trap depths much larger than ${\sim} k_B \times 20 \unit{\mu{}K}$ we observe a strong, density-dependent anomalous heating when the Raman cooling is turned off. 
(Heating rates of up to $10 \unit{\mu{}K/s}$ are observed at \mbox{$n\approx 2{\times}10^{14} \unit{cm^{-3}}$} with a trap depth of $ k_B \times 250 \unit{\mu{}K}$.)
We surmise that at high atomic densities $n \gtrsim 10^{14} \unit{cm^{-3}}$, recombination products of inelastic three-body collisions undergo grazing collisions with trapped atoms, depositing heat in the cloud~\cite{SuppMat}.
This limit on the maximum trap depth is akin to the necessity to maintain, even in the absence of evaporative cooling, a sufficiently low trap depth by applying a so-called ``RF-shield'' in magnetic traps which allows highly energetic nonthermal atoms to escape~\cite{Mewes1996,Burt1997}.

The BEC transition is crossed with $N{\approx 2.5}{\times}10^4 \unit{atoms}$ at a critical temperature of $T_{\mathrm{c}}{=0.61(4)} \unit{\mu K}$.
We are able to reach condensate fractions $N_0/N$ up to $7\%$.
We also verify that if the Raman cooling is turned off during stage X3, the PSD does not increase, and a condensate does not appear.
Furthermore the condensate can be maintained for ${\sim} 1\unit{s}$ after creation if the cooling is left on, but if the cooling is turned off, the condensate decays within ${\sim} 100\unit{ms}$.
This confirms that evaporation is insufficient to create or maintain a condensate in this trap configuration, and that the laser cooling is responsible for inducing the phase transition.

For most laser cooling methods, the requisite spontaneous photon scattering sets a recoil temperature limit. 
Nonetheless, we achieve sub-recoil temperatures by addressing atoms in the high-energy wings of the thermal distribution.
The optical pumping $\ket{2,-1} {\rightarrow} \ket{2,-2}$ requires on average the spontaneous scattering of three photons, and therefore imparts $6\Erecoil$ of energy, where \mbox{$\Erecoil=\hbar^2/(2m \lambdabar^2) = h {\times 3.6} \unit{kHz}$} is the recoil energy of a $795 \unit{nm}$ photon.
As a result, only atoms with $K_{\Deltak}/h \gtrsim 29 \unit{kHz}$ of kinetic energy along the $\Deltak$-direction can be cooled at all~\cite{SuppMat}.
This sets an effective recoil temperature $\Trecoil^{\mathrm{eff}} = 2.8 \unit{\mu K}$. 
We achieve cooling below this effective recoil limit down to $0.5 \unit{\mu{}K}$, i.e. a mean kinetic energy $\langle K_{\Deltak} \rangle /h = 5.2 \unit{kHz}$, by detuning the Raman coupling so that atoms with more than the average kinetic energy are addressed by the cooling light~\cite{SuppMat}.
However, this slows down the cooling for temperatures below  $\Trecoil^{\mathrm{eff}}$ (stage S2 onwards), while inelastic collisions add an increased heat load at high densities.
In X3 when we cross $T_{\mathrm{c}}$, we find that under optimized cooling the Raman transition removes $15 \unit{kHz}$ of kinetic energy, 30\% less than the expected $6\Erecoil$ of heating (see SM~\cite{SuppMat}).
This could indicate that the cooling is aided by bosonic stimulation into the condensate during the photon scattering process, in a similar fashion to Ref.~\cite{Robins2008}.

\label{optimization}

The improved performance of our scheme compared to previous Raman cooling results~\cite{Lee1996,Perrin1999} is primarily due to the flexibility to perform optical pumping to a dark state at large detuning from atomic resonances by operating on the $\mathrm{D_1}$-line.
To identify suitable detunings, we separately characterized light-induced losses over a $16\unit{GHz}$ frequency range to the red of the bare atomic transition, as shown in Fig.~\ref{OptimizationScans}(a), and further detailed in the SM~\cite{SuppMat}.
Fig.~\ref{OptimizationScans}(b) displays the final atom number and condensate fraction of the optimized sequence as a function of detuning around the value of $-4.33 \unit{GHz}$ chosen for the experiment. We find that a suitable detuning has to be maintained within $\pm 50$~MHz to ensure good cooling performance.

Another parameter that needs to be optimized is the photon scattering rate $\Gammasc$ for optical pumping into the $\ket{2,-2}$ dark state. 
Despite the large detuning, the reabsorption of spontaneously scattered optical pumping photons by other atoms is a resonant two-photon process that can lead to excess heating. 
However, it was shown theoretically~\cite{Cirac1996,Castin1998}, and confirmed experimentally~\cite{Wolf2000}, that the excess heating can be suppressed at sufficiently low scattering rate $\Gammasc$, such that the confinement and two-photon Doppler broadening reduce the reabsorption probability.
This limit is known as the \emph{festina lente} regime.
The degradation of the performance at larger $\Gammasc$ in Fig.~\ref{OptimizationScans}(c) is consistent with increased rescattering, as the calculated reabsorption probability approaches unity~\cite{SuppMat}.
A too small value of $\Gammasc$, on the other hand, leads to higher temperatures as parasitic heating mechanisms cannot be compensated when the cooling is too slow.

While $\Rb$ has relatively favorable collision properties (low two-body inelastic loss rate coefficient, and moderate three-body loss rate coefficient in the upper hyperfine manifold), these properties are not unique, and other atomic species may also be suitable for direct laser cooling to BEC.
Since the cooling is not \emph{deeply} subrecoil, relatively high densities $n \sim \lambdabar^{-3} $ are required for reaching BEC.
Thanks to the fast cooling, the effect of inelastic loss is small enough if a cloud is stable at these densities (typ. lifetime $\gtrsim 1\unit{s}$ at $10^{14} \unit{cm^{-3}}$).
Inelastic processes can be further reduced in an effectively one-dimensional geometry \cite{Hu2017}, where fermionization of the bosonic wavefunction reduces collisional processes.
The demonstrated technique could also be directly applied to fermionic atoms~\cite{DeMarco1999}, as well as to laser cooled molecules~\cite{Barry2014,Anderegg2018}. 

\label{conclusion}

In conclusion, we have realized the decades-old goal of BEC purely by laser cooling by creating a single, moderately sized Bose-Einstein condensate in a standard crossed optical dipole trap.
Notably, the method is consistent with the general theoretical recipe put forward by~\citet{Santos2000}.
Further work is needed to explore the limits of this new technique in terms of speed, condensate size and final temperature.
[We observe that the performance is still density-limited, see Fig.~\ref{BEC}(a), and verified that when allowing for moderate final evaporation, nearly pure condensates can be created within a cooling time of $1 \unit{s}$.]
It may also be interesting to investigate if this technique can be used to experimentally realize an atom laser, where the condensate is created by bosonic stimulation into the atomic final state during the spontaneous photon scattering~\cite{Spreeuw1995,Olshanii1996}, rather than through thermalizing elastic collisions.

\begin{acknowledgments}
We would like to thank Cheng Chin, Wolfgang Ketterle, and Martin Zwierlein for stimulating discussions and insightful comments.
A.~A. acknowledges support from the SNSF grant No.~P2ELP2-181926. 
This work was supported by the NSF, NSF CUA, NASA, and MURI through ONR. 
\end{acknowledgments}

\bibliography{BEC_Stuff}

\begin{thebibliography}{43}%
\makeatletter
\providecommand \@ifxundefined [1]{%
 \@ifx{#1\undefined}
}%
\providecommand \@ifnum [1]{%
 \ifnum #1\expandafter \@firstoftwo
 \else \expandafter \@secondoftwo
 \fi
}%
\providecommand \@ifx [1]{%
 \ifx #1\expandafter \@firstoftwo
 \else \expandafter \@secondoftwo
 \fi
}%
\providecommand \natexlab [1]{#1}%
\providecommand \enquote  [1]{``#1''}%
\providecommand \bibnamefont  [1]{#1}%
\providecommand \bibfnamefont [1]{#1}%
\providecommand \citenamefont [1]{#1}%
\providecommand \href@noop [0]{\@secondoftwo}%
\providecommand \href [0]{\begingroup \@sanitize@url \@href}%
\providecommand \@href[1]{\@@startlink{#1}\@@href}%
\providecommand \@@href[1]{\endgroup#1\@@endlink}%
\providecommand \@sanitize@url [0]{\catcode `\\12\catcode `\$12\catcode
  `\&12\catcode `\#12\catcode `\^12\catcode `\_12\catcode `\%12\relax}%
\providecommand \@@startlink[1]{}%
\providecommand \@@endlink[0]{}%
\providecommand \url  [0]{\begingroup\@sanitize@url \@url }%
\providecommand \@url [1]{\endgroup\@href {#1}{\urlprefix }}%
\providecommand \urlprefix  [0]{URL }%
\providecommand \Eprint [0]{\href }%
\providecommand \doibase [0]{http://dx.doi.org/}%
\providecommand \selectlanguage [0]{\@gobble}%
\providecommand \bibinfo  [0]{\@secondoftwo}%
\providecommand \bibfield  [0]{\@secondoftwo}%
\providecommand \translation [1]{[#1]}%
\providecommand \BibitemOpen [0]{}%
\providecommand \bibitemStop [0]{}%
\providecommand \bibitemNoStop [0]{.\EOS\space}%
\providecommand \EOS [0]{\spacefactor3000\relax}%
\providecommand \BibitemShut  [1]{\csname bibitem#1\endcsname}%
\let\auto@bib@innerbib\@empty
\bibitem [{\citenamefont {Bloch}\ \emph {et~al.}(2012)\citenamefont {Bloch},
  \citenamefont {Dalibard},\ and\ \citenamefont
  {Nascimb{\`{e}}ne}}]{Bloch2012}%
  \BibitemOpen
  \bibfield  {author} {\bibinfo {author} {\bibfnamefont {Immanuel}\
  \bibnamefont {Bloch}}, \bibinfo {author} {\bibfnamefont {Jean}\ \bibnamefont
  {Dalibard}}, \ and\ \bibinfo {author} {\bibfnamefont {Sylvain}\ \bibnamefont
  {Nascimb{\`{e}}ne}},\ }\bibfield  {title} {\enquote {\bibinfo {title}
  {{Quantum simulations with ultracold quantum gases}},}\ }\href {\doibase
  10.1038/nphys2259} {\bibfield  {journal} {\bibinfo  {journal} {Nature
  Physics}\ }\textbf {\bibinfo {volume} {8}},\ \bibinfo {pages} {267--276}
  (\bibinfo {year} {2012})}\BibitemShut {NoStop}%
\bibitem [{\citenamefont {Becker}\ \emph {et~al.}(2018)\citenamefont {Becker},
  \citenamefont {Lachmann}, \citenamefont {Seidel}, \citenamefont {Ahlers},
  \citenamefont {Dinkelaker}, \citenamefont {Grosse}, \citenamefont {Hellmig},
  \citenamefont {M{\"{u}}ntinga}, \citenamefont {Schkolnik}, \citenamefont
  {Wendrich}, \citenamefont {Wenzlawski}, \citenamefont {Weps}, \citenamefont
  {Corgier}, \citenamefont {Franz}, \citenamefont {Gaaloul}, \citenamefont
  {Herr}, \citenamefont {L{\"{u}}dtke}, \citenamefont {Popp}, \citenamefont
  {Amri}, \citenamefont {Duncker}, \citenamefont {Erbe}, \citenamefont
  {Kohfeldt}, \citenamefont {Kubelka-Lange}, \citenamefont {Braxmaier},
  \citenamefont {Charron}, \citenamefont {Ertmer}, \citenamefont {Krutzik},
  \citenamefont {L{\"{a}}mmerzahl}, \citenamefont {Peters}, \citenamefont
  {Schleich}, \citenamefont {Sengstock}, \citenamefont {Walser}, \citenamefont
  {Wicht}, \citenamefont {Windpassinger},\ and\ \citenamefont
  {Rasel}}]{Becker2018}%
  \BibitemOpen
  \bibfield  {author} {\bibinfo {author} {\bibfnamefont {Dennis}\ \bibnamefont
  {Becker}}, \bibinfo {author} {\bibfnamefont {Maike~D.}\ \bibnamefont
  {Lachmann}}, \bibinfo {author} {\bibfnamefont {Stephan~T.}\ \bibnamefont
  {Seidel}}, \bibinfo {author} {\bibfnamefont {Holger}\ \bibnamefont {Ahlers}},
  \bibinfo {author} {\bibfnamefont {Aline~N.}\ \bibnamefont {Dinkelaker}},
  \bibinfo {author} {\bibfnamefont {Jens}\ \bibnamefont {Grosse}}, \bibinfo
  {author} {\bibfnamefont {Ortwin}\ \bibnamefont {Hellmig}}, \bibinfo {author}
  {\bibfnamefont {Hauke}\ \bibnamefont {M{\"{u}}ntinga}}, \bibinfo {author}
  {\bibfnamefont {Vladimir}\ \bibnamefont {Schkolnik}}, \bibinfo {author}
  {\bibfnamefont {Thijs}\ \bibnamefont {Wendrich}}, \bibinfo {author}
  {\bibfnamefont {Andr{\'{e}}}\ \bibnamefont {Wenzlawski}}, \bibinfo {author}
  {\bibfnamefont {Benjamin}\ \bibnamefont {Weps}}, \bibinfo {author}
  {\bibfnamefont {Robin}\ \bibnamefont {Corgier}}, \bibinfo {author}
  {\bibfnamefont {Tobias}\ \bibnamefont {Franz}}, \bibinfo {author}
  {\bibfnamefont {Naceur}\ \bibnamefont {Gaaloul}}, \bibinfo {author}
  {\bibfnamefont {Waldemar}\ \bibnamefont {Herr}}, \bibinfo {author}
  {\bibfnamefont {Daniel}\ \bibnamefont {L{\"{u}}dtke}}, \bibinfo {author}
  {\bibfnamefont {Manuel}\ \bibnamefont {Popp}}, \bibinfo {author}
  {\bibfnamefont {Sirine}\ \bibnamefont {Amri}}, \bibinfo {author}
  {\bibfnamefont {Hannes}\ \bibnamefont {Duncker}}, \bibinfo {author}
  {\bibfnamefont {Maik}\ \bibnamefont {Erbe}}, \bibinfo {author} {\bibfnamefont
  {Anja}\ \bibnamefont {Kohfeldt}}, \bibinfo {author} {\bibfnamefont
  {Andr{\'{e}}}\ \bibnamefont {Kubelka-Lange}}, \bibinfo {author}
  {\bibfnamefont {Claus}\ \bibnamefont {Braxmaier}}, \bibinfo {author}
  {\bibfnamefont {Eric}\ \bibnamefont {Charron}}, \bibinfo {author}
  {\bibfnamefont {Wolfgang}\ \bibnamefont {Ertmer}}, \bibinfo {author}
  {\bibfnamefont {Markus}\ \bibnamefont {Krutzik}}, \bibinfo {author}
  {\bibfnamefont {Claus}\ \bibnamefont {L{\"{a}}mmerzahl}}, \bibinfo {author}
  {\bibfnamefont {Achim}\ \bibnamefont {Peters}}, \bibinfo {author}
  {\bibfnamefont {Wolfgang~P.}\ \bibnamefont {Schleich}}, \bibinfo {author}
  {\bibfnamefont {Klaus}\ \bibnamefont {Sengstock}}, \bibinfo {author}
  {\bibfnamefont {Reinhold}\ \bibnamefont {Walser}}, \bibinfo {author}
  {\bibfnamefont {Andreas}\ \bibnamefont {Wicht}}, \bibinfo {author}
  {\bibfnamefont {Patrick}\ \bibnamefont {Windpassinger}}, \ and\ \bibinfo
  {author} {\bibfnamefont {Ernst~M.}\ \bibnamefont {Rasel}},\ }\bibfield
  {title} {\enquote {\bibinfo {title} {{Space-borne Bose–Einstein
  condensation for precision interferometry}},}\ }\href {\doibase
  10.1038/s41586-018-0605-1} {\bibfield  {journal} {\bibinfo  {journal}
  {Nature}\ }\textbf {\bibinfo {volume} {562}},\ \bibinfo {pages} {391--395}
  (\bibinfo {year} {2018})}\BibitemShut {NoStop}%
\bibitem [{\citenamefont {Ketterle}\ and\ \citenamefont
  {Druten}(1996)}]{Ketterle1996a}%
  \BibitemOpen
  \bibfield  {author} {\bibinfo {author} {\bibfnamefont {Wolfgang}\
  \bibnamefont {Ketterle}}\ and\ \bibinfo {author} {\bibfnamefont {N.J.~Van}\
  \bibnamefont {Druten}},\ }\bibfield  {title} {\enquote {\bibinfo {title}
  {{Evaporative Cooling of Trapped Atoms}},}\ }\href {\doibase
  10.1016/S1049-250X(08)60101-9} {\bibfield  {journal} {\bibinfo  {journal}
  {Advances In Atomic, Molecular, and Optical Physics}\ }\textbf {\bibinfo
  {volume} {37}},\ \bibinfo {pages} {181--236} (\bibinfo {year}
  {1996})}\BibitemShut {NoStop}%
\bibitem [{\citenamefont {Anderson}\ \emph {et~al.}(1995)\citenamefont
  {Anderson}, \citenamefont {Ensher}, \citenamefont {Matthews}, \citenamefont
  {Wieman},\ and\ \citenamefont {Cornell}}]{Anderson1995}%
  \BibitemOpen
  \bibfield  {author} {\bibinfo {author} {\bibfnamefont {M.~H.}\ \bibnamefont
  {Anderson}}, \bibinfo {author} {\bibfnamefont {J.~R.}\ \bibnamefont
  {Ensher}}, \bibinfo {author} {\bibfnamefont {M.~R.}\ \bibnamefont
  {Matthews}}, \bibinfo {author} {\bibfnamefont {C.~E.}\ \bibnamefont
  {Wieman}}, \ and\ \bibinfo {author} {\bibfnamefont {E.~A.}\ \bibnamefont
  {Cornell}},\ }\bibfield  {title} {\enquote {\bibinfo {title} {{Observation of
  Bose-Einstein Condensation in a Dilute Atomic Vapor}},}\ }\href {\doibase
  10.1126/science.269.5221.198} {\bibfield  {journal} {\bibinfo  {journal}
  {Science}\ }\textbf {\bibinfo {volume} {269}},\ \bibinfo {pages} {198--201}
  (\bibinfo {year} {1995})}\BibitemShut {NoStop}%
\bibitem [{\citenamefont {Davis}\ \emph {et~al.}(1995)\citenamefont {Davis},
  \citenamefont {Mewes}, \citenamefont {Andrews}, \citenamefont {van Druten},
  \citenamefont {Durfee}, \citenamefont {Kurn},\ and\ \citenamefont
  {Ketterle}}]{Davis1995}%
  \BibitemOpen
  \bibfield  {author} {\bibinfo {author} {\bibfnamefont {K.~B.}\ \bibnamefont
  {Davis}}, \bibinfo {author} {\bibfnamefont {M.~O.}\ \bibnamefont {Mewes}},
  \bibinfo {author} {\bibfnamefont {M.~R.}\ \bibnamefont {Andrews}}, \bibinfo
  {author} {\bibfnamefont {N.~J.}\ \bibnamefont {van Druten}}, \bibinfo
  {author} {\bibfnamefont {D.~S.}\ \bibnamefont {Durfee}}, \bibinfo {author}
  {\bibfnamefont {D.~M.}\ \bibnamefont {Kurn}}, \ and\ \bibinfo {author}
  {\bibfnamefont {W.}~\bibnamefont {Ketterle}},\ }\bibfield  {title} {\enquote
  {\bibinfo {title} {{Bose-Einstein Condensation in a Gas of Sodium Atoms}},}\
  }\href {\doibase 10.1103/PhysRevLett.75.3969} {\bibfield  {journal} {\bibinfo
   {journal} {Phys. Rev. Lett.}\ }\textbf {\bibinfo {volume} {75}},\ \bibinfo
  {pages} {3969--3973} (\bibinfo {year} {1995})}\BibitemShut {NoStop}%
\bibitem [{\citenamefont {DeMarco}\ and\ \citenamefont
  {Jin}(1999)}]{DeMarco1999}%
  \BibitemOpen
  \bibfield  {author} {\bibinfo {author} {\bibfnamefont {B.}~\bibnamefont
  {DeMarco}}\ and\ \bibinfo {author} {\bibfnamefont {D.~S.}\ \bibnamefont
  {Jin}},\ }\bibfield  {title} {\enquote {\bibinfo {title} {{Onset of Fermi
  Degeneracy in a Trapped Atomic Gas}},}\ }\href {\doibase
  10.1126/science.285.5434.1703} {\bibfield  {journal} {\bibinfo  {journal}
  {Science}\ }\textbf {\bibinfo {volume} {285}},\ \bibinfo {pages} {1703--1706}
  (\bibinfo {year} {1999})}\BibitemShut {NoStop}%
\bibitem [{\citenamefont {Barrett}\ \emph {et~al.}(2001)\citenamefont
  {Barrett}, \citenamefont {Sauer},\ and\ \citenamefont
  {Chapman}}]{Barrett2001}%
  \BibitemOpen
  \bibfield  {author} {\bibinfo {author} {\bibfnamefont {M.~D.}\ \bibnamefont
  {Barrett}}, \bibinfo {author} {\bibfnamefont {J.~A.}\ \bibnamefont {Sauer}},
  \ and\ \bibinfo {author} {\bibfnamefont {M.~S.}\ \bibnamefont {Chapman}},\
  }\bibfield  {title} {\enquote {\bibinfo {title} {{All-Optical Formation of an
  Atomic Bose-Einstein Condensate}},}\ }\href {\doibase
  10.1103/PhysRevLett.87.010404} {\bibfield  {journal} {\bibinfo  {journal}
  {Physical Review Letters}\ }\textbf {\bibinfo {volume} {87}},\ \bibinfo
  {pages} {010404} (\bibinfo {year} {2001})}\BibitemShut {NoStop}%
\bibitem [{\citenamefont {Aspect}\ \emph {et~al.}(1988)\citenamefont {Aspect},
  \citenamefont {Arimondo}, \citenamefont {Kaiser}, \citenamefont
  {Vansteenkiste},\ and\ \citenamefont {Cohen-Tannoudji}}]{Aspect1988}%
  \BibitemOpen
  \bibfield  {author} {\bibinfo {author} {\bibfnamefont {A.}~\bibnamefont
  {Aspect}}, \bibinfo {author} {\bibfnamefont {E.}~\bibnamefont {Arimondo}},
  \bibinfo {author} {\bibfnamefont {R.}~\bibnamefont {Kaiser}}, \bibinfo
  {author} {\bibfnamefont {N.}~\bibnamefont {Vansteenkiste}}, \ and\ \bibinfo
  {author} {\bibfnamefont {C.}~\bibnamefont {Cohen-Tannoudji}},\ }\bibfield
  {title} {\enquote {\bibinfo {title} {{Laser Cooling below the One-Photon
  Recoil Energy by Velocity-Selective Coherent Population Trapping}},}\ }\href
  {\doibase 10.1103/PhysRevLett.61.826} {\bibfield  {journal} {\bibinfo
  {journal} {Physical Review Letters}\ }\textbf {\bibinfo {volume} {61}},\
  \bibinfo {pages} {826--829} (\bibinfo {year} {1988})}\BibitemShut {NoStop}%
\bibitem [{\citenamefont {Lee}\ \emph {et~al.}(1996)\citenamefont {Lee},
  \citenamefont {Adams}, \citenamefont {Kasevich},\ and\ \citenamefont
  {Chu}}]{Lee1996}%
  \BibitemOpen
  \bibfield  {author} {\bibinfo {author} {\bibfnamefont {H.~J.}\ \bibnamefont
  {Lee}}, \bibinfo {author} {\bibfnamefont {C.~S.}\ \bibnamefont {Adams}},
  \bibinfo {author} {\bibfnamefont {M.}~\bibnamefont {Kasevich}}, \ and\
  \bibinfo {author} {\bibfnamefont {S.}~\bibnamefont {Chu}},\ }\bibfield
  {title} {\enquote {\bibinfo {title} {{Raman Cooling of Atoms in an Optical
  Dipole Trap}},}\ }\href {\doibase 10.1103/PhysRevLett.76.2658} {\bibfield
  {journal} {\bibinfo  {journal} {Physical Review Letters}\ }\textbf {\bibinfo
  {volume} {76}},\ \bibinfo {pages} {2658--2661} (\bibinfo {year}
  {1996})}\BibitemShut {NoStop}%
\bibitem [{\citenamefont {Kerman}\ \emph {et~al.}(2000)\citenamefont {Kerman},
  \citenamefont {Vuleti{\'{c}}}, \citenamefont {Chin},\ and\ \citenamefont
  {Chu}}]{Kerman2000}%
  \BibitemOpen
  \bibfield  {author} {\bibinfo {author} {\bibfnamefont {Andrew~J.}\
  \bibnamefont {Kerman}}, \bibinfo {author} {\bibfnamefont {Vladan}\
  \bibnamefont {Vuleti{\'{c}}}}, \bibinfo {author} {\bibfnamefont {Cheng}\
  \bibnamefont {Chin}}, \ and\ \bibinfo {author} {\bibfnamefont {Steven}\
  \bibnamefont {Chu}},\ }\bibfield  {title} {\enquote {\bibinfo {title}
  {{Beyond Optical Molasses: 3D Raman Sideband Cooling of Atomic Cesium to High
  Phase-Space Density}},}\ }\href {\doibase 10.1103/PhysRevLett.84.439}
  {\bibfield  {journal} {\bibinfo  {journal} {Physical Review Letters}\
  }\textbf {\bibinfo {volume} {84}},\ \bibinfo {pages} {439--442} (\bibinfo
  {year} {2000})}\BibitemShut {NoStop}%
\bibitem [{\citenamefont {Han}\ \emph {et~al.}(2000)\citenamefont {Han},
  \citenamefont {Wolf}, \citenamefont {Oliver}, \citenamefont {McCormick},
  \citenamefont {DePue},\ and\ \citenamefont {Weiss}}]{Han2000}%
  \BibitemOpen
  \bibfield  {author} {\bibinfo {author} {\bibfnamefont {Dian-Jiun}\
  \bibnamefont {Han}}, \bibinfo {author} {\bibfnamefont {Steffen}\ \bibnamefont
  {Wolf}}, \bibinfo {author} {\bibfnamefont {Steven}\ \bibnamefont {Oliver}},
  \bibinfo {author} {\bibfnamefont {Colin}\ \bibnamefont {McCormick}}, \bibinfo
  {author} {\bibfnamefont {Marshall~T.}\ \bibnamefont {DePue}}, \ and\ \bibinfo
  {author} {\bibfnamefont {David~S.}\ \bibnamefont {Weiss}},\ }\bibfield
  {title} {\enquote {\bibinfo {title} {{3D Raman Sideband Cooling of Cesium
  Atoms at High Density}},}\ }\href {\doibase 10.1103/PhysRevLett.85.724}
  {\bibfield  {journal} {\bibinfo  {journal} {Physical Review Letters}\
  }\textbf {\bibinfo {volume} {85}},\ \bibinfo {pages} {724--727} (\bibinfo
  {year} {2000})}\BibitemShut {NoStop}%
\bibitem [{\citenamefont {Burnett}\ \emph {et~al.}(1996)\citenamefont
  {Burnett}, \citenamefont {Julienne},\ and\ \citenamefont
  {Suominen}}]{Burnett1996}%
  \BibitemOpen
  \bibfield  {author} {\bibinfo {author} {\bibfnamefont {K.}~\bibnamefont
  {Burnett}}, \bibinfo {author} {\bibfnamefont {P.~S.}\ \bibnamefont
  {Julienne}}, \ and\ \bibinfo {author} {\bibfnamefont {K.-A.}\ \bibnamefont
  {Suominen}},\ }\bibfield  {title} {\enquote {\bibinfo {title} {{Laser-Driven
  Collisions between Atoms in a Bose-Einstein Condensed Gas}},}\ }\href
  {\doibase 10.1103/PhysRevLett.77.1416} {\bibfield  {journal} {\bibinfo
  {journal} {Physical Review Letters}\ }\textbf {\bibinfo {volume} {77}},\
  \bibinfo {pages} {1416--1419} (\bibinfo {year} {1996})}\BibitemShut {NoStop}%
\bibitem [{\citenamefont {Ido}\ \emph {et~al.}(2000)\citenamefont {Ido},
  \citenamefont {Isoya},\ and\ \citenamefont {Katori}}]{Ido2000}%
  \BibitemOpen
  \bibfield  {author} {\bibinfo {author} {\bibfnamefont {Tetsuya}\ \bibnamefont
  {Ido}}, \bibinfo {author} {\bibfnamefont {Yoshitomo}\ \bibnamefont {Isoya}},
  \ and\ \bibinfo {author} {\bibfnamefont {Hidetoshi}\ \bibnamefont {Katori}},\
  }\bibfield  {title} {\enquote {\bibinfo {title} {{Optical-dipole trapping of
  Sr atoms at a high phase-space density}},}\ }\href {\doibase
  10.1103/PhysRevA.61.061403} {\bibfield  {journal} {\bibinfo  {journal}
  {Physical Review A}\ }\textbf {\bibinfo {volume} {61}},\ \bibinfo {pages}
  {061403(R)} (\bibinfo {year} {2000})}\BibitemShut {NoStop}%
\bibitem [{\citenamefont {Stellmer}\ \emph {et~al.}(2013)\citenamefont
  {Stellmer}, \citenamefont {Pasquiou}, \citenamefont {Grimm},\ and\
  \citenamefont {Schreck}}]{Stellmer2013}%
  \BibitemOpen
  \bibfield  {author} {\bibinfo {author} {\bibfnamefont {Simon}\ \bibnamefont
  {Stellmer}}, \bibinfo {author} {\bibfnamefont {Benjamin}\ \bibnamefont
  {Pasquiou}}, \bibinfo {author} {\bibfnamefont {Rudolf}\ \bibnamefont
  {Grimm}}, \ and\ \bibinfo {author} {\bibfnamefont {Florian}\ \bibnamefont
  {Schreck}},\ }\bibfield  {title} {\enquote {\bibinfo {title} {{Laser Cooling
  to Quantum Degeneracy}},}\ }\href {\doibase 10.1103/PhysRevLett.110.263003}
  {\bibfield  {journal} {\bibinfo  {journal} {Physical Review Letters}\
  }\textbf {\bibinfo {volume} {110}},\ \bibinfo {pages} {263003} (\bibinfo
  {year} {2013})}\BibitemShut {NoStop}%
\bibitem [{\citenamefont {Hamann}\ \emph {et~al.}(1998)\citenamefont {Hamann},
  \citenamefont {Haycock}, \citenamefont {Klose}, \citenamefont {Pax},
  \citenamefont {Deutsch},\ and\ \citenamefont {Jessen}}]{Hamann1998}%
  \BibitemOpen
  \bibfield  {author} {\bibinfo {author} {\bibfnamefont {S.~E.}\ \bibnamefont
  {Hamann}}, \bibinfo {author} {\bibfnamefont {D.~L.}\ \bibnamefont {Haycock}},
  \bibinfo {author} {\bibfnamefont {G.}~\bibnamefont {Klose}}, \bibinfo
  {author} {\bibfnamefont {P.~H.}\ \bibnamefont {Pax}}, \bibinfo {author}
  {\bibfnamefont {I.~H.}\ \bibnamefont {Deutsch}}, \ and\ \bibinfo {author}
  {\bibfnamefont {P.~S.}\ \bibnamefont {Jessen}},\ }\bibfield  {title}
  {\enquote {\bibinfo {title} {{Resolved-Sideband Raman Cooling to the Ground
  State of an Optical Lattice}},}\ }\href {\doibase
  10.1103/PhysRevLett.80.4149} {\bibfield  {journal} {\bibinfo  {journal}
  {Physical Review Letters}\ }\textbf {\bibinfo {volume} {80}},\ \bibinfo
  {pages} {4149--4152} (\bibinfo {year} {1998})}\BibitemShut {NoStop}%
\bibitem [{\citenamefont {Vuleti{\'{c}}}\ \emph {et~al.}(1998)\citenamefont
  {Vuleti{\'{c}}}, \citenamefont {Chin}, \citenamefont {Kerman},\ and\
  \citenamefont {Chu}}]{Vuletic1998}%
  \BibitemOpen
  \bibfield  {author} {\bibinfo {author} {\bibfnamefont {Vladan}\ \bibnamefont
  {Vuleti{\'{c}}}}, \bibinfo {author} {\bibfnamefont {Cheng}\ \bibnamefont
  {Chin}}, \bibinfo {author} {\bibfnamefont {Andrew~J.}\ \bibnamefont
  {Kerman}}, \ and\ \bibinfo {author} {\bibfnamefont {Steven}\ \bibnamefont
  {Chu}},\ }\bibfield  {title} {\enquote {\bibinfo {title} {{Degenerate Raman
  Sideband Cooling of Trapped Cesium Atoms at Very High Atomic Densities}},}\
  }\href {\doibase 10.1103/PhysRevLett.81.5768} {\bibfield  {journal} {\bibinfo
   {journal} {Physical Review Letters}\ }\textbf {\bibinfo {volume} {81}},\
  \bibinfo {pages} {5768--5771} (\bibinfo {year} {1998})}\BibitemShut {NoStop}%
\bibitem [{\citenamefont {Hu}\ \emph {et~al.}(2017)\citenamefont {Hu},
  \citenamefont {Urvoy}, \citenamefont {Vendeiro}, \citenamefont
  {Cr{\'{e}}pel}, \citenamefont {Chen},\ and\ \citenamefont
  {Vuleti{\'{c}}}}]{Hu2017}%
  \BibitemOpen
  \bibfield  {author} {\bibinfo {author} {\bibfnamefont {Jiazhong}\
  \bibnamefont {Hu}}, \bibinfo {author} {\bibfnamefont {Alban}\ \bibnamefont
  {Urvoy}}, \bibinfo {author} {\bibfnamefont {Zachary}\ \bibnamefont
  {Vendeiro}}, \bibinfo {author} {\bibfnamefont {Valentin}\ \bibnamefont
  {Cr{\'{e}}pel}}, \bibinfo {author} {\bibfnamefont {Wenlan}\ \bibnamefont
  {Chen}}, \ and\ \bibinfo {author} {\bibfnamefont {Vladan}\ \bibnamefont
  {Vuleti{\'{c}}}},\ }\bibfield  {title} {\enquote {\bibinfo {title} {{Creation
  of a Bose-condensed gas of $^{87}$Rb by laser cooling.}}}\ }\href {\doibase
  10.1126/science.aan5614} {\bibfield  {journal} {\bibinfo  {journal} {Science
  (New York, N.Y.)}\ }\textbf {\bibinfo {volume} {358}},\ \bibinfo {pages}
  {1078--1080} (\bibinfo {year} {2017})}\BibitemShut {NoStop}%
\bibitem [{\citenamefont {Kasevich}\ and\ \citenamefont
  {Chu}(1992)}]{Kasevich1992}%
  \BibitemOpen
  \bibfield  {author} {\bibinfo {author} {\bibfnamefont {Mark}\ \bibnamefont
  {Kasevich}}\ and\ \bibinfo {author} {\bibfnamefont {Steven}\ \bibnamefont
  {Chu}},\ }\bibfield  {title} {\enquote {\bibinfo {title} {{Laser cooling
  below a photon recoil with three-level atoms}},}\ }\href {\doibase
  10.1103/PhysRevLett.69.1741} {\bibfield  {journal} {\bibinfo  {journal}
  {Physical Review Letters}\ }\textbf {\bibinfo {volume} {69}},\ \bibinfo
  {pages} {1741--1744} (\bibinfo {year} {1992})}\BibitemShut {NoStop}%
\bibitem [{\citenamefont {Reichel}\ \emph {et~al.}(1995)\citenamefont
  {Reichel}, \citenamefont {Bardou}, \citenamefont {Ben~Dahan}, \citenamefont
  {Peik}, \citenamefont {Rand}, \citenamefont {Salomon},\ and\ \citenamefont
  {Cohen-Tannoudji}}]{Reichel1995}%
  \BibitemOpen
  \bibfield  {author} {\bibinfo {author} {\bibfnamefont {J.}~\bibnamefont
  {Reichel}}, \bibinfo {author} {\bibfnamefont {F.}~\bibnamefont {Bardou}},
  \bibinfo {author} {\bibfnamefont {M.}~\bibnamefont {Ben~Dahan}}, \bibinfo
  {author} {\bibfnamefont {E.}~\bibnamefont {Peik}}, \bibinfo {author}
  {\bibfnamefont {S.}~\bibnamefont {Rand}}, \bibinfo {author} {\bibfnamefont
  {C.}~\bibnamefont {Salomon}}, \ and\ \bibinfo {author} {\bibfnamefont
  {C.}~\bibnamefont {Cohen-Tannoudji}},\ }\bibfield  {title} {\enquote
  {\bibinfo {title} {{Raman Cooling of Cesium below 3 nK: New Approach Inspired
  by L{\'{e}}vy Flight Statistics}},}\ }\href {\doibase
  10.1103/PhysRevLett.75.4575} {\bibfield  {journal} {\bibinfo  {journal}
  {Physical Review Letters}\ }\textbf {\bibinfo {volume} {75}},\ \bibinfo
  {pages} {4575--4578} (\bibinfo {year} {1995})}\BibitemShut {NoStop}%
\bibitem [{\citenamefont {Perrin}\ \emph {et~al.}(1999)\citenamefont {Perrin},
  \citenamefont {Kuhn}, \citenamefont {Bouchoule}, \citenamefont {Pfau},\ and\
  \citenamefont {Salomon}}]{Perrin1999}%
  \BibitemOpen
  \bibfield  {author} {\bibinfo {author} {\bibfnamefont {H}~\bibnamefont
  {Perrin}}, \bibinfo {author} {\bibfnamefont {A}~\bibnamefont {Kuhn}},
  \bibinfo {author} {\bibfnamefont {I}~\bibnamefont {Bouchoule}}, \bibinfo
  {author} {\bibfnamefont {T}~\bibnamefont {Pfau}}, \ and\ \bibinfo {author}
  {\bibfnamefont {C}~\bibnamefont {Salomon}},\ }\bibfield  {title} {\enquote
  {\bibinfo {title} {{Raman cooling of spin-polarized cesium atoms in a crossed
  dipole trap}},}\ }\href {\doibase 10.1209/epl/i1999-00236-0} {\bibfield
  {journal} {\bibinfo  {journal} {Europhysics Letters (EPL)}\ }\textbf
  {\bibinfo {volume} {46}},\ \bibinfo {pages} {141--147} (\bibinfo {year}
  {1999})}\BibitemShut {NoStop}%
\bibitem [{\citenamefont {Boyer}\ \emph {et~al.}(2004)\citenamefont {Boyer},
  \citenamefont {Lising}, \citenamefont {Rolston},\ and\ \citenamefont
  {Phillips}}]{Boyer2004}%
  \BibitemOpen
  \bibfield  {author} {\bibinfo {author} {\bibfnamefont {V.}~\bibnamefont
  {Boyer}}, \bibinfo {author} {\bibfnamefont {L.~J.}\ \bibnamefont {Lising}},
  \bibinfo {author} {\bibfnamefont {S.~L.}\ \bibnamefont {Rolston}}, \ and\
  \bibinfo {author} {\bibfnamefont {W.~D.}\ \bibnamefont {Phillips}},\
  }\bibfield  {title} {\enquote {\bibinfo {title} {{Deeply subrecoil
  two-dimensional Raman cooling}},}\ }\href {\doibase
  10.1103/PhysRevA.70.043405} {\bibfield  {journal} {\bibinfo  {journal} {Phys.
  Rev. A}\ }\textbf {\bibinfo {volume} {70}},\ \bibinfo {pages} {043405}
  (\bibinfo {year} {2004})}\BibitemShut {NoStop}%
\bibitem [{\citenamefont {Hijmans}\ \emph {et~al.}(1996)\citenamefont
  {Hijmans}, \citenamefont {Shlyapnikov},\ and\ \citenamefont
  {Burin}}]{Hijmans1996}%
  \BibitemOpen
  \bibfield  {author} {\bibinfo {author} {\bibfnamefont {T.~W.}\ \bibnamefont
  {Hijmans}}, \bibinfo {author} {\bibfnamefont {G.~V.}\ \bibnamefont
  {Shlyapnikov}}, \ and\ \bibinfo {author} {\bibfnamefont {A.~L.}\ \bibnamefont
  {Burin}},\ }\bibfield  {title} {\enquote {\bibinfo {title} {{Influence of
  radiative interatomic collisions on dark-state cooling}},}\ }\href {\doibase
  10.1103/PhysRevA.54.4332} {\bibfield  {journal} {\bibinfo  {journal}
  {Physical Review A}\ }\textbf {\bibinfo {volume} {54}},\ \bibinfo {pages}
  {4332--4338} (\bibinfo {year} {1996})}\BibitemShut {NoStop}%
\bibitem [{\citenamefont {R{\"{u}}hrig}\ \emph {et~al.}(2015)\citenamefont
  {R{\"{u}}hrig}, \citenamefont {B{\"{a}}uerle}, \citenamefont {Griesmaier},\
  and\ \citenamefont {Pfau}}]{Ruhrig2015}%
  \BibitemOpen
  \bibfield  {author} {\bibinfo {author} {\bibfnamefont {Jahn}\ \bibnamefont
  {R{\"{u}}hrig}}, \bibinfo {author} {\bibfnamefont {Tobias}\ \bibnamefont
  {B{\"{a}}uerle}}, \bibinfo {author} {\bibfnamefont {Axel}\ \bibnamefont
  {Griesmaier}}, \ and\ \bibinfo {author} {\bibfnamefont {Tilman}\ \bibnamefont
  {Pfau}},\ }\bibfield  {title} {\enquote {\bibinfo {title} {{High efficiency
  demagnetization cooling by suppression of light-assisted collisions}},}\
  }\href {\doibase 10.1364/OE.23.005596} {\bibfield  {journal} {\bibinfo
  {journal} {Optics Express}\ }\textbf {\bibinfo {volume} {23}},\ \bibinfo
  {pages} {5596} (\bibinfo {year} {2015})}\BibitemShut {NoStop}%
\bibitem [{\citenamefont {Cirac}\ \emph {et~al.}(1996)\citenamefont {Cirac},
  \citenamefont {Lewenstein},\ and\ \citenamefont {Zoller}}]{Cirac1996}%
  \BibitemOpen
  \bibfield  {author} {\bibinfo {author} {\bibfnamefont {J.~I}\ \bibnamefont
  {Cirac}}, \bibinfo {author} {\bibfnamefont {M}~\bibnamefont {Lewenstein}}, \
  and\ \bibinfo {author} {\bibfnamefont {P}~\bibnamefont {Zoller}},\ }\bibfield
   {title} {\enquote {\bibinfo {title} {{Collective laser cooling of trapped
  atoms}},}\ }\href {\doibase 10.1209/epl/i1996-00165-4} {\bibfield  {journal}
  {\bibinfo  {journal} {Europhysics Letters (EPL)}\ }\textbf {\bibinfo {volume}
  {35}},\ \bibinfo {pages} {647--652} (\bibinfo {year} {1996})}\BibitemShut
  {NoStop}%
\bibitem [{\citenamefont {Castin}\ \emph {et~al.}(1998)\citenamefont {Castin},
  \citenamefont {Cirac},\ and\ \citenamefont {Lewenstein}}]{Castin1998}%
  \BibitemOpen
  \bibfield  {author} {\bibinfo {author} {\bibfnamefont {Y.}~\bibnamefont
  {Castin}}, \bibinfo {author} {\bibfnamefont {J.~I.}\ \bibnamefont {Cirac}}, \
  and\ \bibinfo {author} {\bibfnamefont {M.}~\bibnamefont {Lewenstein}},\
  }\bibfield  {title} {\enquote {\bibinfo {title} {{Reabsorption of Light by
  Trapped Atoms}},}\ }\href {\doibase 10.1103/PhysRevLett.80.5305} {\bibfield
  {journal} {\bibinfo  {journal} {Physical Review Letters}\ }\textbf {\bibinfo
  {volume} {80}},\ \bibinfo {pages} {5305--5308} (\bibinfo {year}
  {1998})}\BibitemShut {NoStop}%
\bibitem [{\citenamefont {Santos}\ \emph {et~al.}(2000)\citenamefont {Santos},
  \citenamefont {Idziaszek}, \citenamefont {Cirac},\ and\ \citenamefont
  {Lewenstein}}]{Santos2000}%
  \BibitemOpen
  \bibfield  {author} {\bibinfo {author} {\bibfnamefont {L}~\bibnamefont
  {Santos}}, \bibinfo {author} {\bibfnamefont {Z}~\bibnamefont {Idziaszek}},
  \bibinfo {author} {\bibfnamefont {J~I}\ \bibnamefont {Cirac}}, \ and\
  \bibinfo {author} {\bibfnamefont {M}~\bibnamefont {Lewenstein}},\ }\bibfield
  {title} {\enquote {\bibinfo {title} {{Laser-induced condensation of trapped
  bosonic gases}},}\ }\href {\doibase 10.1088/0953-4075/33/19/322} {\bibfield
  {journal} {\bibinfo  {journal} {Journal of Physics B: Atomic, Molecular and
  Optical Physics}\ }\textbf {\bibinfo {volume} {33}},\ \bibinfo {pages}
  {4131--4148} (\bibinfo {year} {2000})}\BibitemShut {NoStop}%
\bibitem [{Sup()}]{SuppMat}%
  \BibitemOpen
  \href@noop {} {}\bibinfo {note} {See Supplemental Material at [\emph{URL will
  be inserted by publisher}], which includes
  Ref.~\cite{Burnett1996,Hijmans1996,Jelassi2006a,Jelassi2006,Jelassi2014,Hu2017,Kasevich1992,Lee1996,Perrin1999,Schuster2001,Wolf2017,Cirac1996,Castin1998,Santos2000},
  for details on the optical pumping detuning, the experimental setup, the
  effective recoil limit, the optimization procedure, and the derivation of the
  reabsorption probability.}\BibitemShut {Stop}%
\bibitem [{\citenamefont {Dalfovo}\ \emph {et~al.}(1999)\citenamefont
  {Dalfovo}, \citenamefont {Giorgini}, \citenamefont {Pitaevskii},\ and\
  \citenamefont {Stringari}}]{Dalfovo1999}%
  \BibitemOpen
  \bibfield  {author} {\bibinfo {author} {\bibfnamefont {Franco}\ \bibnamefont
  {Dalfovo}}, \bibinfo {author} {\bibfnamefont {Stefano}\ \bibnamefont
  {Giorgini}}, \bibinfo {author} {\bibfnamefont {Lev~P.}\ \bibnamefont
  {Pitaevskii}}, \ and\ \bibinfo {author} {\bibfnamefont {Sandro}\ \bibnamefont
  {Stringari}},\ }\bibfield  {title} {\enquote {\bibinfo {title} {{Theory of
  Bose-Einstein condensation in trapped gases}},}\ }\href {\doibase
  10.1103/RevModPhys.71.463} {\bibfield  {journal} {\bibinfo  {journal}
  {Reviews of Modern Physics}\ }\textbf {\bibinfo {volume} {71}},\ \bibinfo
  {pages} {463--512} (\bibinfo {year} {1999})}\BibitemShut {NoStop}%
\bibitem [{\citenamefont {Ketterle}\ and\ \citenamefont {van
  Druten}(1996)}]{Ketterle1996}%
  \BibitemOpen
  \bibfield  {author} {\bibinfo {author} {\bibfnamefont {Wolfgang}\
  \bibnamefont {Ketterle}}\ and\ \bibinfo {author} {\bibfnamefont {N.~J.}\
  \bibnamefont {van Druten}},\ }\bibfield  {title} {\enquote {\bibinfo {title}
  {{Bose-Einstein condensation of a finite number of particles trapped in one
  or three dimensions}},}\ }\href {\doibase 10.1103/PhysRevA.54.656} {\bibfield
   {journal} {\bibinfo  {journal} {Physical Review A}\ }\textbf {\bibinfo
  {volume} {54}},\ \bibinfo {pages} {656--660} (\bibinfo {year}
  {1996})}\BibitemShut {NoStop}%
\bibitem [{\citenamefont {Smith}\ \emph {et~al.}(2011)\citenamefont {Smith},
  \citenamefont {Campbell}, \citenamefont {Tammuz},\ and\ \citenamefont
  {Hadzibabic}}]{Smith2011}%
  \BibitemOpen
  \bibfield  {author} {\bibinfo {author} {\bibfnamefont {Robert~P.}\
  \bibnamefont {Smith}}, \bibinfo {author} {\bibfnamefont {Robert L.~D.}\
  \bibnamefont {Campbell}}, \bibinfo {author} {\bibfnamefont {Naaman}\
  \bibnamefont {Tammuz}}, \ and\ \bibinfo {author} {\bibfnamefont {Zoran}\
  \bibnamefont {Hadzibabic}},\ }\bibfield  {title} {\enquote {\bibinfo {title}
  {{Effects of Interactions on the Critical Temperature of a Trapped Bose
  Gas}},}\ }\href {\doibase 10.1103/PhysRevLett.106.250403} {\bibfield
  {journal} {\bibinfo  {journal} {Physical Review Letters}\ }\textbf {\bibinfo
  {volume} {106}},\ \bibinfo {pages} {250403} (\bibinfo {year}
  {2011})}\BibitemShut {NoStop}%
\bibitem [{\citenamefont {Mewes}\ \emph {et~al.}(1996)\citenamefont {Mewes},
  \citenamefont {Andrews}, \citenamefont {van Druten}, \citenamefont {Kurn},
  \citenamefont {Durfee},\ and\ \citenamefont {Ketterle}}]{Mewes1996}%
  \BibitemOpen
  \bibfield  {author} {\bibinfo {author} {\bibfnamefont {M.-O.}\ \bibnamefont
  {Mewes}}, \bibinfo {author} {\bibfnamefont {M.~R.}\ \bibnamefont {Andrews}},
  \bibinfo {author} {\bibfnamefont {N.~J.}\ \bibnamefont {van Druten}},
  \bibinfo {author} {\bibfnamefont {D.~M.}\ \bibnamefont {Kurn}}, \bibinfo
  {author} {\bibfnamefont {D.~S.}\ \bibnamefont {Durfee}}, \ and\ \bibinfo
  {author} {\bibfnamefont {W.}~\bibnamefont {Ketterle}},\ }\bibfield  {title}
  {\enquote {\bibinfo {title} {{Bose-Einstein Condensation in a Tightly
  Confining dc Magnetic Trap}},}\ }\href {\doibase 10.1103/PhysRevLett.77.416}
  {\bibfield  {journal} {\bibinfo  {journal} {Physical Review Letters}\
  }\textbf {\bibinfo {volume} {77}},\ \bibinfo {pages} {416--419} (\bibinfo
  {year} {1996})}\BibitemShut {NoStop}%
\bibitem [{\citenamefont {Burt}\ \emph {et~al.}(1997)\citenamefont {Burt},
  \citenamefont {Ghrist}, \citenamefont {Myatt}, \citenamefont {Holland},
  \citenamefont {Cornell},\ and\ \citenamefont {Wieman}}]{Burt1997}%
  \BibitemOpen
  \bibfield  {author} {\bibinfo {author} {\bibfnamefont {E.~A.}\ \bibnamefont
  {Burt}}, \bibinfo {author} {\bibfnamefont {R.~W.}\ \bibnamefont {Ghrist}},
  \bibinfo {author} {\bibfnamefont {C.~J.}\ \bibnamefont {Myatt}}, \bibinfo
  {author} {\bibfnamefont {M.~J.}\ \bibnamefont {Holland}}, \bibinfo {author}
  {\bibfnamefont {E.~A.}\ \bibnamefont {Cornell}}, \ and\ \bibinfo {author}
  {\bibfnamefont {C.~E.}\ \bibnamefont {Wieman}},\ }\bibfield  {title}
  {\enquote {\bibinfo {title} {{Coherence, Correlations, and Collisions: What
  One Learns about Bose-Einstein Condensates from Their Decay}},}\ }\href
  {\doibase 10.1103/PhysRevLett.79.337} {\bibfield  {journal} {\bibinfo
  {journal} {Physical Review Letters}\ }\textbf {\bibinfo {volume} {79}},\
  \bibinfo {pages} {337--340} (\bibinfo {year} {1997})}\BibitemShut {NoStop}%
\bibitem [{\citenamefont {Robins}\ \emph {et~al.}(2008)\citenamefont {Robins},
  \citenamefont {Figl}, \citenamefont {Jeppesen}, \citenamefont {Dennis},\ and\
  \citenamefont {Close}}]{Robins2008}%
  \BibitemOpen
  \bibfield  {author} {\bibinfo {author} {\bibfnamefont {Nicholas~P.}\
  \bibnamefont {Robins}}, \bibinfo {author} {\bibfnamefont {Cristina}\
  \bibnamefont {Figl}}, \bibinfo {author} {\bibfnamefont {Matthew}\
  \bibnamefont {Jeppesen}}, \bibinfo {author} {\bibfnamefont {Graham~R.}\
  \bibnamefont {Dennis}}, \ and\ \bibinfo {author} {\bibfnamefont {John~D.}\
  \bibnamefont {Close}},\ }\bibfield  {title} {\enquote {\bibinfo {title} {{A
  pumped atom laser}},}\ }\href {\doibase 10.1038/nphys1027} {\bibfield
  {journal} {\bibinfo  {journal} {Nature Physics}\ }\textbf {\bibinfo {volume}
  {4}},\ \bibinfo {pages} {731--736} (\bibinfo {year} {2008})}\BibitemShut
  {NoStop}%
\bibitem [{\citenamefont {Wolf}\ \emph {et~al.}(2000)\citenamefont {Wolf},
  \citenamefont {Oliver},\ and\ \citenamefont {Weiss}}]{Wolf2000}%
  \BibitemOpen
  \bibfield  {author} {\bibinfo {author} {\bibfnamefont {Steffen}\ \bibnamefont
  {Wolf}}, \bibinfo {author} {\bibfnamefont {Steven~J.}\ \bibnamefont
  {Oliver}}, \ and\ \bibinfo {author} {\bibfnamefont {David~S.}\ \bibnamefont
  {Weiss}},\ }\bibfield  {title} {\enquote {\bibinfo {title} {{Suppression of
  Recoil Heating by an Optical Lattice}},}\ }\href {\doibase
  10.1103/PhysRevLett.85.4249} {\bibfield  {journal} {\bibinfo  {journal}
  {Physical Review Letters}\ }\textbf {\bibinfo {volume} {85}},\ \bibinfo
  {pages} {4249--4252} (\bibinfo {year} {2000})}\BibitemShut {NoStop}%
\bibitem [{\citenamefont {Barry}\ \emph {et~al.}(2014)\citenamefont {Barry},
  \citenamefont {McCarron}, \citenamefont {Norrgard}, \citenamefont
  {Steinecker},\ and\ \citenamefont {DeMille}}]{Barry2014}%
  \BibitemOpen
  \bibfield  {author} {\bibinfo {author} {\bibfnamefont {J.~F.}\ \bibnamefont
  {Barry}}, \bibinfo {author} {\bibfnamefont {D.~J.}\ \bibnamefont {McCarron}},
  \bibinfo {author} {\bibfnamefont {E.~B.}\ \bibnamefont {Norrgard}}, \bibinfo
  {author} {\bibfnamefont {M.~H.}\ \bibnamefont {Steinecker}}, \ and\ \bibinfo
  {author} {\bibfnamefont {D.}~\bibnamefont {DeMille}},\ }\bibfield  {title}
  {\enquote {\bibinfo {title} {{Magneto-optical trapping of a diatomic
  molecule}},}\ }\href {\doibase 10.1038/nature13634} {\bibfield  {journal}
  {\bibinfo  {journal} {Nature}\ }\textbf {\bibinfo {volume} {512}},\ \bibinfo
  {pages} {286--289} (\bibinfo {year} {2014})}\BibitemShut {NoStop}%
\bibitem [{\citenamefont {Anderegg}\ \emph {et~al.}(2018)\citenamefont
  {Anderegg}, \citenamefont {Augenbraun}, \citenamefont {Bao}, \citenamefont
  {Burchesky}, \citenamefont {Cheuk}, \citenamefont {Ketterle},\ and\
  \citenamefont {Doyle}}]{Anderegg2018}%
  \BibitemOpen
  \bibfield  {author} {\bibinfo {author} {\bibfnamefont {Lo{\"{i}}c}\
  \bibnamefont {Anderegg}}, \bibinfo {author} {\bibfnamefont {Benjamin~L.}\
  \bibnamefont {Augenbraun}}, \bibinfo {author} {\bibfnamefont {Yicheng}\
  \bibnamefont {Bao}}, \bibinfo {author} {\bibfnamefont {Sean}\ \bibnamefont
  {Burchesky}}, \bibinfo {author} {\bibfnamefont {Lawrence~W.}\ \bibnamefont
  {Cheuk}}, \bibinfo {author} {\bibfnamefont {Wolfgang}\ \bibnamefont
  {Ketterle}}, \ and\ \bibinfo {author} {\bibfnamefont {John~M.}\ \bibnamefont
  {Doyle}},\ }\bibfield  {title} {\enquote {\bibinfo {title} {{Laser cooling of
  optically trapped molecules}},}\ }\href {\doibase 10.1038/s41567-018-0191-z}
  {\bibfield  {journal} {\bibinfo  {journal} {Nature Physics}\ }\textbf
  {\bibinfo {volume} {14}},\ \bibinfo {pages} {890--893} (\bibinfo {year}
  {2018})}\BibitemShut {NoStop}%
\bibitem [{\citenamefont {Spreeuw}\ \emph {et~al.}(1995)\citenamefont
  {Spreeuw}, \citenamefont {Pfau}, \citenamefont {Janicke},\ and\ \citenamefont
  {Wilkens}}]{Spreeuw1995}%
  \BibitemOpen
  \bibfield  {author} {\bibinfo {author} {\bibfnamefont {R.~J.~C}\ \bibnamefont
  {Spreeuw}}, \bibinfo {author} {\bibfnamefont {T}~\bibnamefont {Pfau}},
  \bibinfo {author} {\bibfnamefont {U}~\bibnamefont {Janicke}}, \ and\ \bibinfo
  {author} {\bibfnamefont {M}~\bibnamefont {Wilkens}},\ }\bibfield  {title}
  {\enquote {\bibinfo {title} {{Laser-like Scheme for Atomic-Matter Waves}},}\
  }\href {\doibase 10.1209/0295-5075/32/6/002} {\bibfield  {journal} {\bibinfo
  {journal} {Europhysics Letters (EPL)}\ }\textbf {\bibinfo {volume} {32}},\
  \bibinfo {pages} {469--474} (\bibinfo {year} {1995})}\BibitemShut {NoStop}%
\bibitem [{\citenamefont {Olshanii}\ \emph {et~al.}(1996)\citenamefont
  {Olshanii}, \citenamefont {Castin},\ and\ \citenamefont
  {Dalibard}}]{Olshanii1996}%
  \BibitemOpen
  \bibfield  {author} {\bibinfo {author} {\bibfnamefont {Maxim}\ \bibnamefont
  {Olshanii}}, \bibinfo {author} {\bibfnamefont {Yvan}\ \bibnamefont {Castin}},
  \ and\ \bibinfo {author} {\bibfnamefont {Jean}\ \bibnamefont {Dalibard}},\
  }\bibfield  {title} {\enquote {\bibinfo {title} {{A model for an atom
  laser}},}\ }in\ \href@noop {} {\emph {\bibinfo {booktitle} {Proc. XII Conf.
  on Laser Spectroscopy}}},\ \bibinfo {editor} {edited by\ \bibinfo {editor}
  {\bibfnamefont {Inguscio}\ \bibnamefont {M.}}, \bibinfo {editor}
  {\bibfnamefont {Allegrini}\ \bibnamefont {M.}}, \ and\ \bibinfo {editor}
  {\bibfnamefont {Sasso}\ \bibnamefont {A.}}}\ (\bibinfo  {publisher} {World
  Scientific},\ \bibinfo {year} {1996})\ pp.\ \bibinfo {pages}
  {7--12}\BibitemShut {NoStop}%
\bibitem [{\citenamefont {Jelassi}\ \emph
  {et~al.}(2006{\natexlab{a}})\citenamefont {Jelassi}, \citenamefont {{Viaris
  de Lesegno}},\ and\ \citenamefont {Pruvost}}]{Jelassi2006a}%
  \BibitemOpen
  \bibfield  {author} {\bibinfo {author} {\bibfnamefont {H.}~\bibnamefont
  {Jelassi}}, \bibinfo {author} {\bibfnamefont {B.}~\bibnamefont {{Viaris de
  Lesegno}}}, \ and\ \bibinfo {author} {\bibfnamefont {L.}~\bibnamefont
  {Pruvost}},\ }\bibfield  {title} {\enquote {\bibinfo {title}
  {{Photoassociation spectroscopy of $^{87}$Rb$_2$ ($5s_{1 / 2} + 5p _{1/2}$)
  $0g^-$ long-range molecular states: Analysis by Lu-Fano graph and improved
  LeRoy-Bernstein formula}},}\ }\href {\doibase 10.1103/PhysRevA.73.032501}
  {\bibfield  {journal} {\bibinfo  {journal} {Physical Review A}\ }\textbf
  {\bibinfo {volume} {73}},\ \bibinfo {pages} {032501} (\bibinfo {year}
  {2006}{\natexlab{a}})}\BibitemShut {NoStop}%
\bibitem [{\citenamefont {Jelassi}\ \emph
  {et~al.}(2006{\natexlab{b}})\citenamefont {Jelassi}, \citenamefont
  {de~Lesegno},\ and\ \citenamefont {Pruvost}}]{Jelassi2006}%
  \BibitemOpen
  \bibfield  {author} {\bibinfo {author} {\bibfnamefont {H.}~\bibnamefont
  {Jelassi}}, \bibinfo {author} {\bibfnamefont {B.~Viaris}\ \bibnamefont
  {de~Lesegno}}, \ and\ \bibinfo {author} {\bibfnamefont {L.}~\bibnamefont
  {Pruvost}},\ }\bibfield  {title} {\enquote {\bibinfo {title}
  {{Photoassociation spectroscopy of $^{87}$Rb$_2$ ($5s_{1 / 2} + 5p _{1/2}$)
  $0u^+$ long-range molecular states: Coupling with the ($5s_{1 / 2} + 5p
  _{3/2}$) $0u^+$ series analyzed using the Lu-Fano approach}},}\ }\href
  {\doibase 10.1103/PhysRevA.74.012510} {\bibfield  {journal} {\bibinfo
  {journal} {Physical Review A}\ }\textbf {\bibinfo {volume} {74}},\ \bibinfo
  {pages} {012510} (\bibinfo {year} {2006}{\natexlab{b}})}\BibitemShut
  {NoStop}%
\bibitem [{\citenamefont {Jelassi}\ and\ \citenamefont
  {Pruvost}(2014)}]{Jelassi2014}%
  \BibitemOpen
  \bibfield  {author} {\bibinfo {author} {\bibfnamefont {Haikel}\ \bibnamefont
  {Jelassi}}\ and\ \bibinfo {author} {\bibfnamefont {Laurence}\ \bibnamefont
  {Pruvost}},\ }\bibfield  {title} {\enquote {\bibinfo {title} {{Weakly bound
  $^{87}$Rb$_2$ ($5s_{1 / 2} + 5p _{1/2}$) $1g$ molecules: Hyperfine
  interaction and LeRoy-Bernstein analysis including linear and nonlinear
  terms}},}\ }\href {\doibase 10.1103/PhysRevA.89.032514} {\bibfield  {journal}
  {\bibinfo  {journal} {Physical Review A}\ }\textbf {\bibinfo {volume} {89}},\
  \bibinfo {pages} {032514} (\bibinfo {year} {2014})}\BibitemShut {NoStop}%
\bibitem [{\citenamefont {Schuster}\ \emph {et~al.}(2001)\citenamefont
  {Schuster}, \citenamefont {Marte}, \citenamefont {Amtage}, \citenamefont
  {Sang}, \citenamefont {Rempe},\ and\ \citenamefont
  {Beijerinck}}]{Schuster2001}%
  \BibitemOpen
  \bibfield  {author} {\bibinfo {author} {\bibfnamefont {J.}~\bibnamefont
  {Schuster}}, \bibinfo {author} {\bibfnamefont {A.}~\bibnamefont {Marte}},
  \bibinfo {author} {\bibfnamefont {S.}~\bibnamefont {Amtage}}, \bibinfo
  {author} {\bibfnamefont {B.}~\bibnamefont {Sang}}, \bibinfo {author}
  {\bibfnamefont {G.}~\bibnamefont {Rempe}}, \ and\ \bibinfo {author}
  {\bibfnamefont {H.~C.~W.}\ \bibnamefont {Beijerinck}},\ }\bibfield  {title}
  {\enquote {\bibinfo {title} {{Avalanches in a Bose-Einstein Condensate}},}\
  }\href {\doibase 10.1103/PhysRevLett.87.170404} {\bibfield  {journal}
  {\bibinfo  {journal} {Physical Review Letters}\ }\textbf {\bibinfo {volume}
  {87}},\ \bibinfo {pages} {170404} (\bibinfo {year} {2001})}\BibitemShut
  {NoStop}%
\bibitem [{\citenamefont {Wolf}\ \emph {et~al.}(2017)\citenamefont {Wolf},
  \citenamefont {Dei{\ss}}, \citenamefont {Kr{\"{u}}kow}, \citenamefont
  {Tiemann}, \citenamefont {Ruzic}, \citenamefont {Wang}, \citenamefont
  {D'Incao}, \citenamefont {Julienne},\ and\ \citenamefont
  {Denschlag}}]{Wolf2017}%
  \BibitemOpen
  \bibfield  {author} {\bibinfo {author} {\bibfnamefont {Joschka}\ \bibnamefont
  {Wolf}}, \bibinfo {author} {\bibfnamefont {Markus}\ \bibnamefont {Dei{\ss}}},
  \bibinfo {author} {\bibfnamefont {Artjom}\ \bibnamefont {Kr{\"{u}}kow}},
  \bibinfo {author} {\bibfnamefont {Eberhard}\ \bibnamefont {Tiemann}},
  \bibinfo {author} {\bibfnamefont {Brandon~P}\ \bibnamefont {Ruzic}}, \bibinfo
  {author} {\bibfnamefont {Yujun}\ \bibnamefont {Wang}}, \bibinfo {author}
  {\bibfnamefont {Jos{\'{e}}~P}\ \bibnamefont {D'Incao}}, \bibinfo {author}
  {\bibfnamefont {Paul~S}\ \bibnamefont {Julienne}}, \ and\ \bibinfo {author}
  {\bibfnamefont {Johannes~Hecker}\ \bibnamefont {Denschlag}},\ }\bibfield
  {title} {\enquote {\bibinfo {title} {{State-to-state chemistry for three-body
  recombination in an ultracold rubidium gas.}}}\ }\href {\doibase
  10.1126/science.aan8721} {\bibfield  {journal} {\bibinfo  {journal} {Science
  (New York, N.Y.)}\ }\textbf {\bibinfo {volume} {358}},\ \bibinfo {pages}
  {921--924} (\bibinfo {year} {2017})}\BibitemShut {NoStop}%
\end{thebibliography}%

\clearpage
\beginsupplement
\widetext

\begin{center}
\textbf{\large Supplemental Materials:\\{}Direct laser cooling to Bose-Einstein condensation in a dipole trap}
\end{center}

\section{Optical Pumping Detuning}

\begin{figure}[h]
\includegraphics[width=16cm]{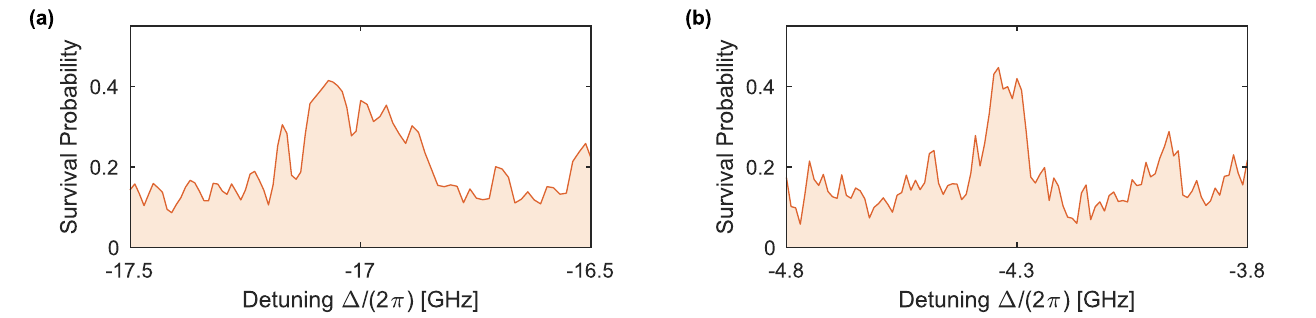}%
\caption{\label{supDetuningScan}
Survival probability of the atoms in the trap as a function of detuning after repeated cycles of optical pumping, corresponding to the scattering of about 100 photons, magnified on two exemplary detuning ranges.
Fig.~\ref{supDetuningScan}(b) is centered around the value used for most of the cooling data presented in the main text.
}
\end{figure}

To find an optical pumping detuning that avoids losses from molecular resonances as predicted in Refs.~\cite{Burnett1996,Hijmans1996}, we start with a cold cloud of atoms and measure the loss spectrum below threshold (i.e. bare atomic resonance) for light-atom detunings $\Delta$ above $-18 \unit{GHz}$.
This is closer to resonance than previously explored experimentally on the $\mathrm{D_1}$-line of $\Rb$~\cite{Jelassi2006a,Jelassi2006,Jelassi2014}.
After executing a sequence which scatters the same number of photons, about 100, regardless of detuning, we obtain a fraction of surviving atoms as given in Fig.~\ref{OptimizationScans}(a) of the main text, with frequency regions where the remaining atom number is relatively large shown in Fig.~\ref{supDetuningScan}.
The scan is performed with a resolution of $10 \unit{MHz}$.
The optimum around $-4.33 \unit{GHz}$ in Fig.~\ref{supDetuningScan}(b) was chosen due to its good performance.
Other optima could be used as well, and better optima may exist outside the range of this scan.

\section{Experimental details}

\begin{figure}[t]
\includegraphics[width=17cm]{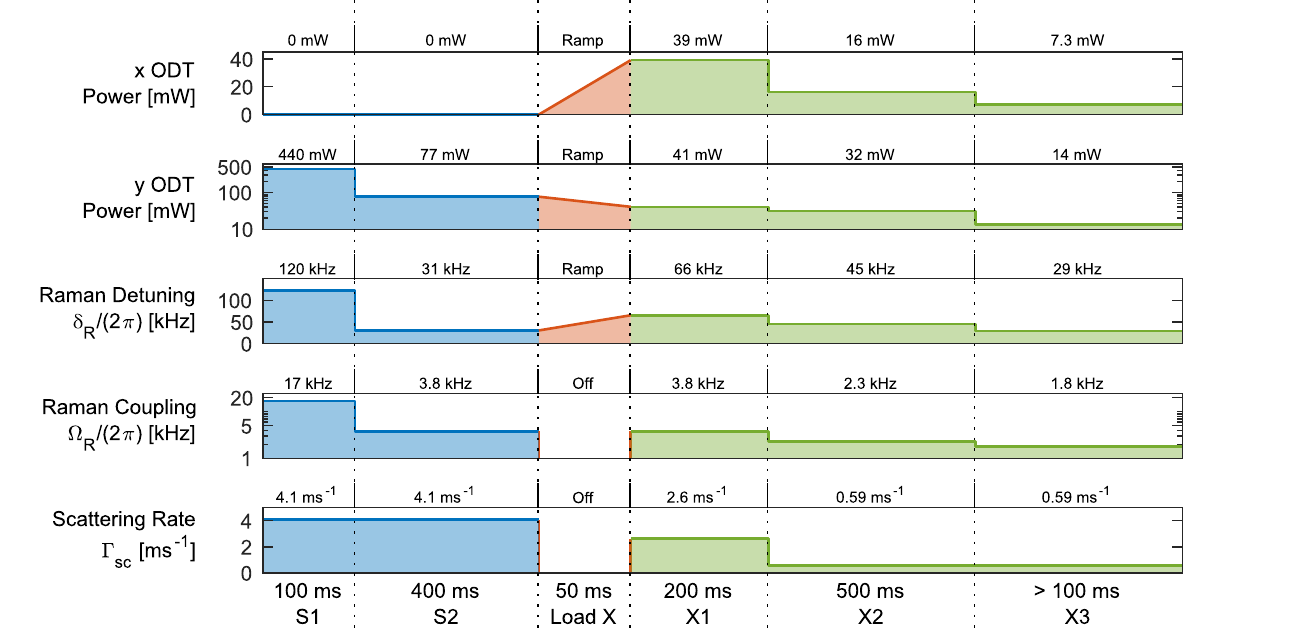}%
\caption{\label{supparamstable}
Optimized multistage Raman cooling sequence.
Initially we cool in single beam ODT (blue), then ramp up a second trapping beam (red) to load into a crossed ODT for the final stages of cooling (green).
}
\end{figure}

The cloud is prepared in a similar way as in Ref.~\cite{Hu2017}. Rb atoms are loaded into a MOT from a thermal vapor, followed by a compressed MOT stage where the optical pumping power is strongly reduced, such that the atoms occupy the $F=1$ ground state manifold.
The ODT is turned on at all times and about $1{\times} 10^5 \unit{atoms}$ are loaded into it when the MOT fields are switched off.

The ODT propagating in the $y$-direction (see Fig.~\ref{fig1} in the main text), in which the atoms are originally loaded, is focused to a waist of about $10 \mum$, while the second ODT propagating in the $x$-direction has a waist of about $18 \mum$ at the position of the atoms.
To avoid interference, the beams differ in frequency by $160\unit{MHz}$.
The powers of each trapping beam throughout the sequence are shown in Fig.~\ref{supparamstable}.
The (calculated) total trap depths, excluding the influence of gravity but including counter-rotating terms, are $430 \unit{\mu{}K}$ in stage S1 and $14 \unit{\mu{}K}$ in stage X3.

The imaging axis is the same as that of the $\pi$ beam, which is slightly rotated by $\approx 17^{\circ}$ from the $x$-axis in the $x-y$ plane. 
In Fig.~\ref{BEC}(b)-(c), we denote the horizontal axis of the image as $y^{\prime}$, which is rotated from the $y$-axis in the $x-y$ plane by the same angle. 

The $\sigmaminus$ optical pumping beam at $795 \unit{nm}$ has a highly elliptical shape at the position of the atoms, with waists of $30 \mum$ along the $x$-direction and $\sim 1\unit{mm}$ along the $y$-direction, to optimally address atoms along the sODT.
The pumping beam creates a sizable light shift $\delta_{\mathrm{LS}}$ of the $\ket{2,-1}$ state, given by $ \delta_{\mathrm{LS}} / \Gammasc = \Delta / \Gamma $, where $\Gamma$ is the natural linewidth of the $5\mathrm{P}_{1/2}$ excited state.
Since the state $\ket{2,-2}$ is dark for the $\sigmaminus$ light, there is no appreciable light shift on this state. 
The light shift is at its largest in stage S1, reaching $ \delta_{\mathrm{LS}}/(2\pi) = 500 \unit{kHz} $.
It is determined experimentally by measuring the shift in the Raman resonance.
From the measured values of $ \delta_{\mathrm{LS}}$ we deduce the scattering rates $\Gammasc$ for $\ket{2,-1}$, shown in Fig.~\ref{supparamstable}.
Since the detuning $\Delta$ is comparable to the hyperfine splitting of the ground state, the same light also pumps atoms out of the $\ket{5\mathrm{S}_{1/2}; F{=}1}$ manifold.

The $\pi$ Raman beam propagates in the $x-y$ plane (see Fig.~\ref{fig1} in the main text), with a waist of $0.5\unit{mm}$.
Its light is derived from the same laser that generates the $\sigmaminus$ beam, but it is detuned by $2\unit{MHz}$ from the $\sigmaminus$ using acousto-optic modulators to avoid interference.
This makes laser frequency noise common mode between the two beams, thereby loosening the requirements on laser linewidth necessary to have a narrow Raman transition.
The Raman beam follows the same path as the light used for absorption imaging, and therefore it is circularly polarized.
Over the course of the cooling only its $\pi$-polarized component (polarization along the $z$-axis, i.e. half its power), contributes to the Raman coupling.
The other half of the power only adds a negligible amount of scattering and light shift to the $\ket{2,-2}$ state.
Very little power ($\lesssim 100 \unit{\mu{}W}$) is required to obtain significant Raman coupling (several kHz, see Fig.~\ref{supparamstable}), so the light shift and scattering rate induced on the $\ket{2,-2}$ state are limited to $\lesssim 0.3 \unit{kHz}$ and $\lesssim 2 \unit{s^{-1}}$, respectively, even at the largest powers used in the cooling sequence.

For all the data presented, each data point is evaluated as an average of 3 to 5 time-of-flight absorption images.

\section{Effective recoil limit}

At trap frequencies in the range $0.1 {-} 5 \unit{kHz}$, smaller than the Raman coupling, the cooling operates mostly in the free space limit with unresolved motional sidebands and outside the Lamb-Dicke regime \cite{Kasevich1992,Lee1996,Perrin1999}.
Each Raman transition from $\ket{2,-2}$ to $\ket{2,-1}$ transfers $- \hbar \Deltak$ of momentum to the atoms, where $\Deltak$ is the difference between the wave vectors of the $\sigmaminus$ and $\pi$ photons.
The kinetic energy removed during one Raman transition for an atom of initial velocity $ \mathbf{v} $ is:
\begin{align}
	\Delta K_\mathrm{Raman} 	&= \hbar \Deltak \cdot \mathbf{v} - \frac{\hbar^2 |\Deltak|^2}{2m} \nonumber \\
				&= \hbar \deltav - 2 \Erecoil
\end{align}
where $\deltav = \Deltak \cdot \mathbf{v}$ is the atom's two-photon Doppler shift.
The orthogonality of the $\pi$ and $\sigmaminus$ beams results in $|\Deltak|^2 = 2 |\lambdabar|^{-2}$, and $\Erecoil = \hbar^2 / (2m \lambdabar^{2})$ is the recoil energy of a $795 \unit{nm}$ photon.
The branching ratio of the optical pumping to the dark state $\ket{2,-2}$ is $1/3$, which results in the scattering of an average of 3 $\sigmaminus$ photons to optically pump the atoms following a Raman transition.
In the limit of low scattering rate ($\Gammasc \ll \omega_{x,y,z}$, which is required to limit reabsorption heating, see below), only the recoil energies of the scattered $\sigmaminus$ photons add, on top of those of the spontaneously emitted photons, resulting in an average net recoil heating of $6 \Erecoil$ per optical pumping cycle.
The average net energy removed per cooling cycle is then given by:
\begin{align}
	\Delta K_\mathrm{total} 	&= \Delta K_\mathrm{Raman} - 6 \Erecoil \nonumber \\
							&= \hbar \deltav - 8 \Erecoil
\end{align}
Therefore there is net cooling if and only if the two-photon Doppler shift is larger than $8 \Erecoil$, which translates to a kinetic energy in the $\Deltak$ direction of:
\begin{align}
	K_{\Deltak}
	&\geq 8 \Erecoil
	= h {\times} 29 \unit{kHz}
\end{align}
As a result, for temperatures below $\Trecoil^{\mathrm{eff}} = 2.8 \unit{\mu{}K}$ ($\langle K_{\Deltak} \rangle = \frac{1}{2} \kB \Trecoil^{\mathrm{eff}}$ along $\Deltak$), the cooling speed drops since only a small fraction of the atoms have Doppler shifts above $8 \Erecoil$ where they can be cooled.

\section{Optimization of the cooling and limiting factors}

\begin{figure}[t]
\includegraphics[width=17.8cm]{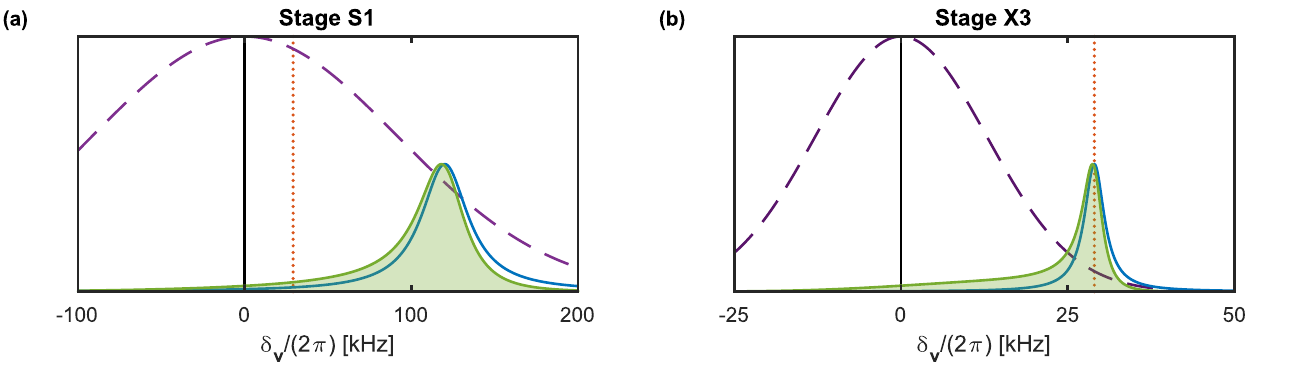}%
\caption{\label{supprofiles}
Thermal velocity distribution (dashed purple), time averaged Lorentzian Raman excitation profile expected given the Raman Rabi frequency (solid blue line), rescaled product of the thermal distribution and the excitation profile (shaded green), and effective recoil limit of $29 \unit{kHz}$ (dotted red) for cooling sequence parameters in (a) stage S1 and (b) stage X3. On average, a cooling cycle will cool atoms with velocity greater than the effective recoil limit, and heat atoms with velocity below the effective recoil limit.
}
\end{figure}

For each stage the Raman transitions are tuned to a particular velocity class by adjusting the magnetic field, which differentially shifts the $\ket{2,-2}$ and $\ket{2,-1}$ states.
In stage S1, where the mean Doppler shift is much larger than the $8 \Erecoil$ limit, we find as expected that the optimal Raman detuning is near the rms Doppler shift, see Fig.~\ref{supprofiles}(a).
This is a compromise between removing a large amount of energy  per cooling cycle ($\Delta K_\mathrm{total} \approx \hbar \deltaRaman $) and having a sufficiently large probability of finding atoms at those velocities (which drops as $\deltaRaman$ increases).
In stage X3 the optimization yields an optimal resonant Doppler shift of $\deltaRaman/(2\pi) = 29 \unit{kHz}$.
As shown in Fig.~\ref{supprofiles}(b), the probability of finding an atom at these velocities is very small, and the amount of energy removed $\Delta K_\mathrm{total}$ also becomes very small.
By averaging $\Delta K_\mathrm{Raman}$ over the actual distribution of addressed atoms [green shaded in Fig.~\ref{supprofiles}(b)] in stage X3, we obtain $\langle \Delta K_\mathrm{Raman} \rangle /h = 15 \unit{kHz}$, well below the recoil heating of $6 \Erecoil /h = 22 \unit{kHz}$, which should lead to heating.
Yet cooling is observed, which suggests bosonic enhancement of the branching ratio into the $\ket{2,-2}$ state due to the emerging condensate.
A better branching ratio would reduce the average recoil heating during optical pumping, and so cooling could be achieved even when the Raman transition removes less than $6 \Erecoil$ of kinetic energy.

At each stage, the strength of the Raman coupling $\OmegaR$ is optimized by scanning the power of the $\pi$ beam.
Too small $\OmegaR$ lead to a narrow excitation profile, and therefore only a small fraction of the atoms undergo a Raman transition.
However, if $\OmegaR$ is too large, already ``cold'' atoms undergo a Raman transition, due to the broadened excitation profile, and are heated during the optical pumping.

Initially in the sODT it is favorable to have a fairly large scattering rate for cooling speed.
The smaller number of atoms available to cool in the Boltzmann tail above the recoil energy in the later stages results in a lower optimal scattering rate as it becomes more favorable to decrease reabsorption heating as seen in Fig.~\ref{OptimizationScans}(c) and discussed below.

For the ODT powers, the main considerations are finding a balance between low density in order to limit inelastic loss and heating, and maintaining large enough trapping frequencies and therefore critical temperatures.
Additionally in the final stage, having a low trap depth has proven crucial to avoid an observed density-dependent heating that increases with trap depth.
As pointed out in Ref.~\cite{Schuster2001}, the products of three-body recombination of $\Rb$ atoms in the $\ket{2,-2}$ state to the least-bound molecular state ($h {\times} 24 \unit{MHz}$ of binding energy~\cite{Wolf2017}) can collide with the cold sample before they leave the trap. This was shown to lead to large loss for a collisionally opaque ensemble.
In our case, the sample is not collisionally opaque (collision probability $\sim 0.1$ for $s$-wave scattering only), nor is the trap deep enough to directly hold recombination products which would dump their energy into the cloud.
The presence of a $d$-wave shape resonance at the energy of the least-bound state for a $\Rb_2$ molecule is expected to enhance the collisional cross-section of recombination products with the trapped atoms, which could lead to strong losses in the collisionally thick regime.
However, we mostly observe heating, which we suspect arises from recombination products undergoing grazing collisions with trapped atoms, with the latter remaining trapped and depositing heat into the cloud.
We found that lowering the trap power as much as possible during the final stage X3 to minimize heating was necessary to reach condensation.

The optimized values of the relevant parameters throughout the sequence, namely trap power, Raman detuning $\deltaRaman$, Raman coupling $\OmegaR$ and scattering rate $\Gammasc$, are shown in Fig.~\ref{supparamstable} for reference.

\section{Estimation of the reabsorption probability}

The reabsorption of a scattered photon causes excess recoil heating. This is especially of concern since the reabsorption cross-section, corresponding to a two-photon resonant process, can take on its maximum possible value $\sim 6 \pi \lambdabar^2$.
Several strategies have been laid out for suppressing this~\cite{Cirac1996,Castin1998,Santos2000}, boiling down to the use of a low scattering rate $\Gammasc$ for optical pumping, a regime known as \emph{festina lente}.
With our optimal parameters, we have $\Gammasc \ll \omega_{\mathrm{D}},\omega_{x,y,z}$, leading to a suppression of the reabsorption process on the order of $\Gammasc / \omega_{\mathrm{D}}$, where $\omega_{\mathrm{D}} $ represents the Doppler width.
The reabsorption cross-section $\sigma_{\mathrm{reabs}}$ is given by~\cite{Castin1998}:
\begin{equation}
	\sigma_{\mathrm{reabs}} \approx 4\pi \lambdabar^2 \frac{\sqrt{\pi} \Gammasc}{2\sqrt{2}\omega_{\mathrm{D}}}
\end{equation}
and the reabsorption probability $p_{\mathrm{reabs}}$ is given by:
\begin{equation}
	p_{\mathrm{reabs}} \approx \sigma_{\mathrm{reabs}} \langle nl \rangle
\end{equation}
where $\langle nl \rangle$ is the mean column density of the cloud.
We use the following formula for the mean column density of a thermal cloud, taken from~\cite{Schuster2001}:
\begin{equation}
	\langle nl \rangle = \sqrt{\frac{\pi}{8}} n(0) \sigma_z \frac{\mathrm{artanh}[\sqrt{1-1/\varepsilon^2}]}{\sqrt{\varepsilon^2-1}}
\end{equation}
where $n(0)$ is the peak density, $\sigma_z$ is the cloud's waist along the short axis of an elongated trap ($z$-axis here), and $\varepsilon=\omega_y/\omega_z$ is the aspect ratio of the trap.
Assuming a classical cloud at the condensation point and a scattering rate $\Gammasc = 0.59 \unit{ms^{-1}}$, we obtain that a probability for reabsorption $p_{\mathrm{reabs}} \sim 0.1$.
Hence reabsorption is not expected to be a significant factor, but it could explain why the performance deteriorates at larger $\Gammasc$ in Fig.~\ref{OptimizationScans}(c) of the main text, since $p_{\mathrm{reabs}}$ varies linearly with $\Gammasc$.

\end{document}